\documentclass[aps,PRX,twocolumn,superscriptaddress,longbibliography]{revtex4-1}

\usepackage[colorlinks=true, allcolors=blue]{hyperref}
\usepackage{upgreek} 
\usepackage{natbib}
\usepackage{graphicx}
\usepackage{ulem}
\usepackage{comment}

\begin{document}

\title{All-optical Differential Radii in Zinc}

\author{B. K. Sahoo}\email{bijaya@prl.res.in}
 \affiliation{
Atomic, Molecular and Optical Physics Division, Physical Research Laboratory, Navrangpura, Ahmedabad 380058, Gujarat, India
}

\author{B. Ohayon}\email{bohayon@technion.ac.il}
\affiliation{
The Helen Diller Quantum Center, Department of Physics,
Technion-Israel Institute of Technology, Haifa, 3200003, Israel
}

\date{\today}

\begin{abstract}
We conduct high-accuracy calculations of isotope shift (IS) factors of the states involving the $D_1$ and $D_2$ lines in Zn II.
Together with a global fit to the available optical IS data, we extract nuclear-model-independent, precise differential radii for a long chain of Zn isotopes. These radii are compared with the ones inferred from muonic X-ray measurements. Some deviations are found, which we ascribe to the deformed nature of Zn nuclei that introduces nuclear-model dependency into radii extractions from muonic atoms.
We arrive at the conclusion that in cases where the many-body atomic calculations of IS factors are well-established, optical determinations of differential radii are more reliable than those extracted from the muonic X-ray measurements, opening the door to improved determination of nuclear radii across the nuclear chart.
\end{abstract}

\maketitle

\section{Introduction}

Isotope shifts (ISs) are the changes in energies of electrons bound to isotopes of the same element. They are sensitive probes of changes in nuclear size and mass. Masses are measured with sufficient accuracy in penning traps~\cite{2018-Traps}, leaving ISs to probe differential root-mean-square (RMS) charge radii, $\delta r^2$.
Optical IS measurements are fast and efficient, and so can be applied to systems with short-lived nuclei, yielding $\delta r^2$ for long chains of isotopes and isomers. These are then used in a variety of nuclear structure investigations extending from proton to neutron drip lines~\cite{2023-Review}. 

$\delta r^2$ may also be evaluated by taking the difference between the absolute radii of two stable isotopes as determined via muonic X-ray spectroscopy~\cite{2004-FH}.
A direct comparison of the $\delta r^2$ obtained from electronic and muonic atoms is not only an important test of state-of-the-art atomic and nuclear theories, but also a powerful vehicle to search for lepton-neutron interactions carried by massive new bosons~\cite{2017-Yotam,2022-g,2022-Clara}.

Extracting $\delta r^2$ from optical ISs is not straightforward, as one needs to estimate the response of the system to changes in mass, the mass shift (MS), and in size, the field shift (FS), with high accuracy;
a demanding task in multielectron systems.
In light systems ($Z\le30$), the MS dominates the IS, and so a high accuracy in its estimation is required in order to be sensitive to $\delta r^2$.

At the precision level relevant to this work, ISs for a particular atomic transition $i$, between isotopes with mass numbers $A$ and $A'$, can be written as 
\begin{eqnarray}
\delta \nu^{A,A'}_i &=& 
K_i\mu^{A,A'} + F_i   (\delta r^2)^{A,A'},
\label{eq:IS}
\end{eqnarray}
where $\mu^{A,A'}=1/M_{A}-1/M_{A'}$ is the inverse nuclear mass difference, and $K_i$ and $F_i$ are the transition-dependent MS and FS factors.
The MS factor can further be divided into normal MS factor $K_i^\mathrm{NMS}$, which is one-body part of the MS operator, and the specific MS factor $K_i^\mathrm{SMS}$, which is its two-body part~\cite{1987-Palmer}.
In order to extract precise values of $\delta r^2$ from measurements of $\delta \nu$ for a set of isotopes, it is imperative to estimate $K_i$ and $F_i$ reliably.

In many cases, including Zn, the transitions for which the IS factors may be calculated precisely and reliably are not the ones that are most useful for measurements with short-lived isotopes \cite{2012-Co}. 
It is thus advantageous to project the calculated factors from one transition to the other. Solving Eq.~(\ref{eq:IS}) for two transitions $i$ and $j$ results in the linear equation
\begin{equation}
\delta\bar{\nu}^{A,A'}_i = 
K_{ij}  + F_{ij}  \delta \bar{\nu}^{A,A'}_j,
\label{eq:lin}
\end{equation}
with $\delta\bar{\nu}^{A,A'}_i \equiv \delta\nu^{A,A'}_i/\mu^{A,A'}$ the reduced ISs, $F_{ij}=F_i/F_j$, and $K_{ij}=K_i-F_{ij}K_j$.
These fitted coefficients may then be utilized to make the aforementioned projection from one transition to another.

In this work, we combine state-of-the-art calculations of IS factors in Zn II with a global fit to available measurements, and extract $\delta r^2$ for a long chain of Zn nuclei. We then show that, contrary to popular belief, this optical determination is not only more precise than one based on muonic atoms; it is also more accurate due to a much reduced dependence on the nuclear model.

\begin{figure*}[t]
\centering
\includegraphics[width=0.95\textwidth]{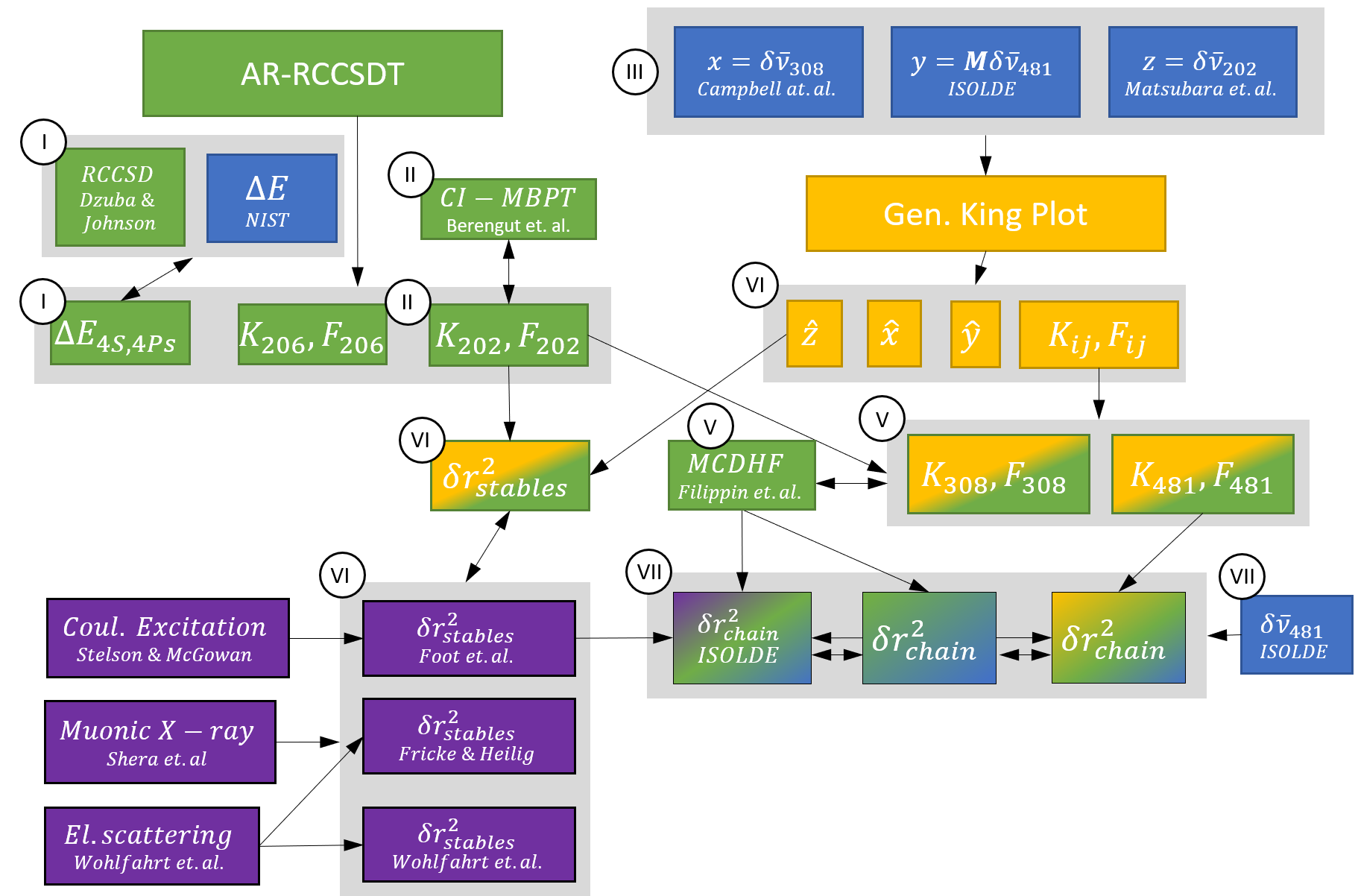}
\caption{
Schematic diagram demonstrating the flow of the present work. Single-ended arrows represent direction from input to output, double-ended arrows denote comparison between results obtained via different methods, Roman numerals indicate the number of the table in which the results are listed, and the box colors denote different groups of data, with gradient-field boxes indicating a combination of groups. Among the box colors, blue means optical measurements, orange stands for global fit and their outputs, green denotes \textit{ab intio} calculations while purple represents measurements in muonic atoms and their inputs. The indicated references are -- Campbell  \textit{et al.}~\cite{1997-308}, ISOLDE~\cite{2019-ISOLDE}, Matsubara \textit{et al.}~\cite{2003-Trap}, Dzuba \& Johnson~\cite{2007-Dzuba},
NIST~\cite{1995-Levels, 2000-Lines}, Berengut \textit{et al.}~\cite{2003-CI}, Stelson \& McGowan~\cite{1962-SM},
Shera \textit{et al.}~\cite{1976-Shera}, Wohlfahrt \textit{et al.}~\cite{1980-MuE}, Foot \textit{et al.}~\cite{1982-Deform}, and Fricke \& Heilig~\cite{2004-FH}.
}
\end{figure*}\label{fig:flow}

The structure of this work is portrayed in Fig.~\ref{fig:flow}.
We first briefly describe the method of calculating energies and IS factors of the first three low-lying states of Zn~II. The results are given in Tables \ref{tab:E} and \ref{tab:factors} at different approximations of the calculation. They are discussed and compared with available data, calculations with another method, and our previous calculations in Cd II~\cite{2022-Cd}.
We then review the relevant experimental IS data, given in Table~\ref{tab:inputs}, and analyze them using a self-consistent global fitting procedure, whose output is given in Table~\ref{tab:outputs}.
Results of the analysis are combined with the calculated factors to produce both precise and accurate $\delta r^2$ for stable isotopes of Zn, given in Table~\ref{tab:rad}. They are compared with the determinations via muonic atom measurements.
A different combination of our calculation and analysis gives IS factors for transitions in Zn I (Table~\ref{tab:ISproj}). These are used to benchmark \textit{ab initio} calculations in Zn I, and to extract $\delta r^2$ for a long chain of short-lived nuclei from measurements done at ISOLDE\,\cite{2019-ISOLDE}. 
The results are given in Table~\ref{tab:Rad2}.

\section{First Principles calculations for Zn II}

\subsection{Method of calculation}

Among the commonly known many-body methods, coupled-cluster (CC) theory is popularly known as the gold standard. This is owing to its ability to account for electron correlation effects competently at a given level of computational requirement, while obeying properties such as size-extensivity and size-consistency. This method is widely applied to atomic, nuclear, molecular, and solid state systems to determine spectroscopic properties meticulously \cite{bartlett,crawford,ccbook,bishop}. We employed the CC method in the relativistic framework (RCC method) by considering the Dirac-Coulomb-Breit (DCB) atomic Hamiltonian ($H$). Corrections from the lower-order quantum electrodynamics (QED) effects are also accounted for using effective model potentials~\cite{bijaya1,bijaya2}. For conveniently obtaining the ground state ($4s ~ ^2S_{1/2}$) and the first two excited states ($4p ~ ^2P_{1/2;3/2}$) with minimum computational cost, we first determine the wave function ($|\Psi_0 \rangle$) of the closed-core ($[3d^{10}]$) of Zn II. The states of interest are then obtained by appending the corresponding valence electron to the $[3d^{10}]$ configuration.
  
The (R)CC theory ansatz is given by~\cite{bartlett,crawford,ccbook,bishop}
\begin{eqnarray}
    |\Psi_0 \rangle = e^{S_0} |\Phi_0 \rangle ,
\end{eqnarray}
where $|\Phi_0 \rangle$ is the mean-field wave function for the $[3d^{10}]$ configuration determined using the Dirac-Hartree-Fock (DHF) method and $S_0$ denotes the excitation RCC operator that is responsible for generating all possible types of configuration state functions acting over $|\Phi_0 \rangle$. The exact energy of the closed-core can be evaluated by \cite{bartlett}
\begin{eqnarray}
E_0 = \frac{\langle \Phi_0 | e^{S_0^{\dagger}} H e^{S_0}   | \Phi_0 \rangle} {\langle \Phi_0 | e^{S_0^{\dagger}} e^{S_0}   | \Phi_0 \rangle} =\langle \Phi_0 | \left ( H e^{S_0} \right )_l  | \Phi_0 \rangle , \label{eqeng0} 
\end{eqnarray}
where subscript $l$ denotes linked terms only. The right-hand side of the above equation follows the condition 
\begin{eqnarray}
\langle \Phi_0^* |  \left ( H e^{S_0} \right )_l | \Phi_0 \rangle = 0 ,
\end{eqnarray}
where $| \Phi_0^* \rangle $ represents all possible excited state determinants with respect to $| \Phi_0 \rangle$. The above equation is adopted to determine amplitudes of the $S_0$ operator. Then, the wave function ($|\Psi_v \rangle$) of a state with the $[3d^{10}]$ closed-core and a valence orbital $v$ is given by
\begin{eqnarray}
    |\Psi_v \rangle = e^{S_v} \left [ a_v^{\dagger} |\Psi_0 \rangle \right ],
\end{eqnarray}
where $S_v$ accounts for all possible electron correlation effects from $|\Psi_0 \rangle$ including the electron from the valence orbital $v$. Due to only one $v$ electron, we can write 
\begin{eqnarray}
    |\Psi_v \rangle = \left \{ 1+ S_v \right  \} e^{S_0} |\Phi_v \rangle ,
\end{eqnarray}
where $|\Phi_v \rangle = a_v^{\dagger} |\Phi_0 \rangle$ is the modified DHF wave function. The amplitude solving equation for $S_v$ yields
\begin{eqnarray}
 \langle \Phi_v^* | \left \{ \left ( H e^{S_0} \right )_l -E_v) \right \} S_v  + \left ( H e^{S_0} \right )_l | \Phi_v \rangle = 0 . \label{eqamp}
\end{eqnarray}
In the above expression, $| \Phi_v^* \rangle $ denotes for all possible excited state determinants with respect to $| \Phi_v \rangle$. The energy of the respective state is given by
\begin{eqnarray}
 E_v = \langle \Phi_v | \left ( H e^{S_0} \right )_l \{ 1+ S_v \} | \Phi_v \rangle .\label{eqeng}
\end{eqnarray}
We note that Eqs.~(\ref{eqamp}) and~(\ref{eqeng}) are coupled, so they form non-linear equations. The expressions in Eqs.~(\ref{eqeng0}) and~(\ref{eqeng}) contain only a finite number of terms and can be easily evaluated. Difference between these two energy values gives us the electron affinity (EA) for the $v$ electron.
We have considered electron correlation effects first from the single and double excitations in the RCC theory (RCCSD method) up to 20$s$, 20$p$, 19$d$, 18$f$ and 16$g$ orbitals. Since considering triple excitations among all these orbitals was not feasible with the available computational resources, we have allowed triple excitations up to 18$s$, 18$p$, 17$d$ and 11$f$ orbitals to account for the correlation effects along with the singles and doubles excitations among the aforementioned orbitals (RCCSDT method).

The IS factors can be determined by solving the following first-order perturbed equation 
\begin{eqnarray}
  (H-E_v^{(0)}) [ \lambda |\Psi_v^{(1)} \rangle ] &=& \lambda (O_\mathrm{IS} -  H_\mathrm{IS} ) |\Psi_v^{(0)} \rangle ,
\end{eqnarray}
where $H_\mathrm{IS}$ is the IS Hamiltonian for the respective IS factor $O_\mathrm{IS}$. The superscripts $(0)$ and $(1)$ denote contributions from the DCB Hamiltonian $H$ and first-order correction (identified by $\lambda=1$) due to $H_\mathrm{IS}$, respectively. 

In the analytical response RCC theory (AR-RCC method), we can evaluate $O_\mathrm{IS}$ by expanding the RCC operators as
\begin{eqnarray}
    S_0 = S_0^{(0)} + \lambda S_0^{(1)} \ \ \ \text{and} \ \ \ S_v = S_v^{(0)} + \lambda S_v^{(1)} .
\end{eqnarray}
Unperturbed wave functions and energies can be obtained using Eqs.~(\ref{eqamp}) and~(\ref{eqeng}), while amplitude-solving equations for the first-order excitation operators in the AR-RCC approach are given by  
\begin{eqnarray}
&& \langle \Phi_0^* |  \left ( H e^{S_0^{(0)}} S_0^{(1)} + H_\mathrm{IS} e^{S_0^{(0)}} \right )_l  | \Phi_0 \rangle = 0 \\
& \text{and} & \nonumber \\
&&  \langle \Phi_v^* | \left \{ \left ( H e^{S_0^{(0)}} \right )_l -E_v^{(0)})  \right \} S_v^{(1)}  + \left ( H e^{S_0} S_0^{(1)} \right )_l \nonumber \\ &&   \times  \left \{ 1+ S_v^{(0)} \right \} + \left ( H_\mathrm{IS} e^{S_0} \right )_l \left \{ 1+ S_v^{(0)} \right \} \nonumber \\ && +O_\mathrm{IS} S_v^{(0)}  | \Phi_v \rangle = 0 .  \ \ \ \ \ \label{eqamp1}
\end{eqnarray}
In the above equation, the expression for an IS factor is given by
\begin{eqnarray}
 O_\mathrm{IS} &=& \langle \Phi_v | \left ( H e^{S_0^{(0)}} \right )_l  S_v^{(1)}  + \left ( H e^{S_0} S_0^{(1)} \right )_l \nonumber \\ && \ \ \ \ \ \ \ \ \ \ \ \ \ \ \ \ \ \ \ \times  \left \{ 1+ S_v^{(0)} \right \} | \Phi_v \rangle . \label{eqeng1}
\end{eqnarray}
Like Eqs.~(\ref{eqamp}) and~(\ref{eqeng}), Eqs.~(\ref{eqamp1}) and~(\ref{eqeng1}) are coupled. This form of an IS factor expression contains all terminating series in the RCC theory framework, so the results are expected to be quite accurate. Especially, it can determine the SMS factors more reliably for which the corresponding $H_\mathrm{IS}$ is a two-body operator. 

\begin{table*}[tb]
\caption{
Calculated EAs (in cm$^{-1}$) of the considered states in $^{64}$Zn at different levels of approximation. The estimated EEs and FS from the EAs are also quoted in the table. Our final results are compared with the experimental values (Exp.) and previously reported theoretical results. Differences between our calculated and experimental values are shown as $\Delta$ in percentage.
}
\begin{ruledtabular}
\begin{tabular}{l rrrrrrr rrr r}
State  & \multicolumn{1}{c}{DHF}  & \multicolumn{1}{c}{MP2}  & \multicolumn{1}{c}{RCCSD} & RCCSDT & $+$Breit & $+$QED & Recoil & Total & Exp.~\cite{1995-Levels, 2000-Lines} & $\Delta$(\%) & Ref.~\cite{2007-Dzuba}\\       
\hline \\ 
$4s ~ ^2S_{1/2}$    & $135134$  & 143606  & $143866$ & $145029$ & $-57$ & $-25$    & $-2$ & $144946$ & $144893(2)~$ & 0.04 & $145334$ \\
$4p ~ ^2P_{1/2}$    &   $90524$  & 95169   & $95612$  & $96499$  & $-50$ & $~~~3$   & $-1$ & $96451$  & $96412(2)~$  & 0.04 & $96613$\\
$4p ~ ^2P_{3/2}$    & $89787$  & 94292   &  $94732$  & $95609$  & $-33$ & $~-3$    & $-1$ & $95572$  & $95538(2)~$  & 0.04 & $95728$\\
$D_1$   &  $44610$    & 48437   & $48254$  & $48530$  & $-7$  & $-27$   & $-1$ & $48495$  & $48481(0.)$ & 0.03 & $48721$\\
$D_2$   &  $45347$  &  49314  & $49133$  & $49419$  & $-24$   & $-21$    & $-1$ & $49374$  & $49355(0.)$ & 0.04 & $49606$\\
FS &  $737$ & 877    & $879$    & $889$    & $-16$ & $ ~~~ 6$ & $~~~0.$ & $879$ & $874(0.)$   & 0.57 & $885$
\end{tabular}
\end{ruledtabular}
\label{tab:E}
\end{table*}

\subsection{Results and discussion: Energies}

To test accuracy of the wave functions using which the IS factors are calculated, we compare differences in EAs, corresponding to excitation energies (EEs), and the fine-structure (FS) splitting, with their measured values. Our results are given in Table~\ref{tab:E} at different approximations in the level of electron excitations and relativistic effects; detailed descriptions regarding these approximations can be found in Ref.~\cite{2021-Li}. The calculations are performed by assuming an infinitely heavy nucleus. To account for the finite nuclear mass, we calculate the first-order recoil corrections, choosing the mass of the most abundant isotope $^{64}$Zn, based on the calculated total MS factors given in Table~\ref{tab:factors}. They are found to be negligibly small.
To test the numerical stability, we carry out the energy calculations by taking $c\rightarrow\infty$ ($\sim 1000$~a.u.) and observe that the EAs of the $^2P_{1/2}$ and $^2P_{3/2}$ states match with each other up to $5$ parts-per-million.

In light multielectron systems, and in contrast to highly charged ions, electron correlation effects play the decisive role for accurate determination of atomic properties. To check convergence in the results with respect to electron correlation effects, we have evaluated EAs using the DHF, second-order relativistic perturbation theory (MP2), RCCSD, and RCCSDT methods systematically. EEs of the $D_1$ and $D_2$ lines, and the FS splitting between the first two excited states are also estimated using the calculated EAs.

As can be seen from Table~\ref{tab:E}, the DHF method gives the lowest EA values. Electron correlation effects gradually increase them from MP2 to RCCSDT through RCCSD method. From the differences between the RCCSD and RCCSDT results, we find that contributions from triple excitations are of the order of $1\%$. This is twice as large as for the corresponding EAs in Cd II~\cite{2022-Cd}, and improves the agreement with experiment from $0.7\%$ to $0.04\%$. Contributions from the Breit, QED and nuclear recoil effects are found to be smaller. Thus, we anticipate that inclusion of contributions from the higher level excitations, such as quadrupole excitations, may improve the results further.

The correlation trends in EEs are found to be slightly different, where MP2 predicts larger values than the RCCSD method. Here we find that the absolute accuracy is increased from $53~$cm$^{-1}$ for the ground level, to $\sim16~$cm$^{-1}$ for the $D_1$ and $D_2$ intervals, indicating that unaccounted-for higher-order effects partly cancel out. Considering the $D_1$ line, the effect of the Breit correction is nearly canceled, leaving only $-7~$cm$^{-1}$, which also compares well with $-5~$cm$^{-1}$ calculated in Ref. \cite{2007-Breit}. The deviation from experiment is $14~$cm$^{-1}$, half that of the approximate QED contribution, which is $-27~$cm$^{-1}$. This makes the $D_1$ transition suitable for benchmarking with refined treatments of QED effects beyond the model potential used in this work. For the $D_2$ transition, both the Breit and QED corrections seem to be equally important with respect to deviation from experiment. Having validated the approximated QED correction with the $D_1$ result, the EE value for the $D_2$ line can be used to test the role of the Breit interaction.
In this case too, we find a comparable contribution from the Breit interaction between our $-24\,$cm$^{-1}$ and the $-20\,$cm$^{-1}$ calculated in Ref.~\cite{2007-Breit}.
Further comparing FS of the $4p$ level with its experimental value, the agreement is within $5\,$cm$^{-1}$.

To our knowledge, this work comprises one of the most accurate calculations of EAs in a singly charged many-electron ion. 
It stems from a balanced treatment of electron correlation effects and higher-order relativistic corrections. 

\subsection{Results and discussion: IS factors}

\begin{table*}[t]
\caption{Calculated IS factors of the ground and the first two excited states of Zn II.
$D_1$ and $D_2$ denote the intervals $^2S_{1/2}-$$^2P_{1/2}$ and $^2S_{1/2}-$$^2P_{3/2}$, respectively. 
NMS values given from the scaling law are obtained using the experimental energies.} 

\begin{ruledtabular}
\begin{tabular}{l rrr rr r r}
State  & \multicolumn{1}{c}{DHF}  & \multicolumn{1}{c}{AR-RCCSD} & AR-RCCSDT & $+$Breit & $+$QED & Total & Other \\

\hline \\ 
\multicolumn{1}{l}{\underline{$F$ MHz/fm$^2$}}&&&&&&& Ref.~\cite{2003-CI}\\
$4s ~ ^2S_{1/2}$ & $-1221$ &  $-1575$ & $-1564(20)$& $4 $ & $19(5)$ & $-1541(21)$   \\
$4p ~ ^2P_{1/2}$ & $-5$    &  $-28$   & $-44(01)$   & $0.$ & $0.$    & $-44(01)$  \\
$4p ~ ^2P_{3/2}$ & $-0.$   &  $-23$   & $-39(02)$    & $0.$ & $0.$    & $-39(02)$ \\
$D_1$            & $-1216$ &  $-1547$ & $-1520(20)$& $3$  & $18(5)$ & $-1498(20)$ & $-1596(80)$\\
$D_2$            & $-1221$ &  $-1552$ & $-1525(20)$  & $3$  & $19(5)$ & $-1503(20)$ &$-1596(80)$\\
\hline \\
\multicolumn{1}{l}{\underline{$K^\mathrm{SMS}$ GHz~u}} &&&&&&& Ref.~\cite{2003-CI}\\
$4s ~ ^2S_{1/2}$ &  $-1576$ & $1654$ & $1679(20)$  & $4$ & $0.$  & $1684(20)$ &  \\
$4p ~ ^2P_{1/2}$ &  $-1176$ & $358 $ & $ 435(15)$  & $1$ & $-1$  & $436(15)$  & \\
$4p ~ ^2P_{3/2}$ &  $-1121$ & $379$ & $ 414(10)$  & $2$ & $-0.$ & $415(10)$  & \\
$D_1$            &  $-400$  & $1297$ & $1244(25)$  & $3$ & $1$   & $1248(25)$ & $1310(69)$\\
$D_2$            &  $-456$  & $1275$ & $1265(22)$  & $3$ & $1$   & $1267(22)$ & $1267(69)$\\
\hline \\ 
\multicolumn{1}{l}{\underline{$K^\mathrm{NMS}$ GHz~u}} &&&&&&& Scaling law\\
$4s ~ ^2S_{1/2}$ &  $4898$ & $2168$ &  $2215(30)$  & $-2$ &  $-1$ & $2211(30)$ &  $2383$ \\
$4p ~ ^2P_{1/2}$ &  $2866$ & $1409$ &  $1478(15)$  & $-3$ &  $0.$ &  $1476(15)$ & $1586$ \\
$4p ~ ^2P_{3/2}$ &  $2815$ & $1393$ &  $1460(15)$  & $-2$ & $-0.$ & $1458(15)$ & $1571$ \\
$D_1$            &  $2031$ & $759$ &  $737(10)$   & $1$  &  $-2$ & $736(10)$ & $797$\\
$D_2$            &  $2083$ & $775$ &  $755(10)$   & $0.$ &  $-1$ & $753(10)$ & $812$\\
\hline \\
\multicolumn{1}{l}{\underline{$K$ GHz~u}} \\
$4s ~ ^2S_{1/2}$ & $3321$ & $3823$ & $3894(36)$  & $2$   & $-1$  & $3895(36)$ & \\
$4p ~ ^2P_{1/2}$ & $1690$ & $1767$ & $1913(21)$  & $-1$  & $-0.$ & $1912(21)$  & \\
$4p ~ ^2P_{3/2}$ & $1694$ & $1851$ & $1874(18)$  & $-0.$ & $-1$  & $1873(18)$  & \\
$D_1$            & $1631$ & $2056$ & $1981(27)$  & $3$   & $-1$  & $1983(27)$    & \\
$D_2$            & $1627$ & $2051$ & $2020(24)$  & $2$   & $-1$  & $2021(24)$    & 
\end{tabular}
\end{ruledtabular}
\label{tab:factors}
\end{table*}

In Table~\ref{tab:factors}, we give the IS factors evaluated using the DHF method and AR-RCC theory at the RCCSD (AR-RCCSD) and RCCSDT (AR-RCCSDT) approximations. 
In the same table, we also quote corrections due to Breit and QED interactions from the AR-RCCSD method, and the results of previous calculations that are reported in Ref.~\cite{2003-CI}. 

The FS in light systems is dominated by the $S$ states for which the electric wave functions have the most overlap with the nucleus. 
Accordingly, at the mean field level, $F_{4s}$ takes a large value, with electron correlation effects amplifying it significantly by $\sim 22\%$. 
The DHF value for the $4P_{1/2}$ state is quite small while it is negligible for the $4P_{3/2}$ state.
However, electron correlation effects make these values noticeable. A similar trend was observed in Cd~II~\cite{2022-Cd}.
However, in contrast with Cd~II, the correlation trends in the ground and excited states are found to be different -- in the ground state  result decrease from the AR-RCCSD method to the AR-RCCSDT method, while in the excited states they go the other way around.
Moreover, triple excitations contribute three times larger in Zn~II than in Cd~II. This highlights the importance of including higher-level excitations for accurate determination of $F$ values in the lighter systems.
Before applying higher-order corrections, we may compare our values with the results from Ref.~\cite{2003-CI}, produced using the combined configuration interaction and many-body perturbation theory (CI$+$MBPT method). The results agree within the larger estimated uncertainty of the CI$+$MBPT calculations.

From the analyses of higher-order relativistic effects, we find the Breit contribution to be negligible while the QED contribution is important on the level of $1\%$, half the contribution in Cd~II. Due to the approximate form of the potential used to estimate the QED contributions, we ascribe to it an uncertainty of $25\%$. Nevertheless, contrary to the situation in Cd~II, this uncertainty is negligible compared with the total quoted uncertainty stemming from unaccounted-for higher level electron excitations, which is in turn negligible as compared with that of the MS, discussed next.

Among all the IS factors, determination of SMS factors are more challenging owing to the two-body form of the SMS interaction operator. Moreover, triple excitations due to the SMS operator contain second-order perturbation contributions. This is why some of the potential many-body methods fail to produce SMS factors precisely. It is evident from Table~\ref{tab:factors} that the signs of the SMS factors from the DHF method and AR-RCC theory are opposite which support the above statement. 
Though it appears that the difference between the SMS factors of the ground state from the AR-RCCSD and AR-RCCSDT methods is small, this is the result of large cancellations among correlation effects arising through the triple excitations. However, such cancellations are not pronounced in the excited states making the differences among the AR-RCCSD and AR-RCCSDT results quite large. We also notice an interesting trend here that the SMS factor of the ground state is much larger than the first $P_{1/2,3/2}$ excited states, similarly to Na I~\cite{2022-Na} and Cd II~\cite{2022-Cd}, but different from Mg II~\cite{2022-Na} and Li-like systems~\cite{2021-Li}, in which the SMS factors of the $P$-states are large. In Ca II, the SMS factors of the $S$ and $P$ level contributions are comparable~\cite{2021-Ca}.
This clearly indicates that even for a relatively simple system with one valence electron, it is difficult to make a guess or estimate from a scaling law the SMS factors without performing rigorous calculations. 
We find the SMS factors of the $D_1$ and $D_2$ lines to agree with the values reported using the CI$+$MBPT method~\cite{2003-CI}, within their larger uncertainties.

Though the NMS operator is a one-body operator, theoretical studies on the NMS factors have been given less attention in the literature. This is owing to two reasons. First, electron correlations associated with the NMS operator are peculiar in nature. Second, it is trivial to estimate them by scaling the experimental energies as per the Viral theorem. However, the theorem relies on the assumption that the system is spherically symmetric. Though atomic systems are so, electron density distributions in different states are not always so. Thus, the  theorem may be more applicable to the $S$-states. Further, the inclusion of relativistic effects may deviate from the Virial theorem assumption. In view of this, it would be interesting to probe the roles of electron correlation effects in the determination of NMS factors and compare with the scaled experimental values.
In Table~\ref{tab:factors} we give $K^\mathrm{NMS}$ calculated from the one body part of the relativistic MS operator at different approximations. None of these values are close to the scaled values, also given in the table.
There are large differences among the values obtained using the DHF method and AR-RCC theory. Differences among the AR-RCCSD and AR-RCCSDT values are noticeable, but not large enough to assert that missing contributions from the higher-level excitations are responsible for the deviation from the scaling law. It is worth mentioning here that the calculated energies, given in Table~\ref{tab:E}, match with their experimental values much better than the estimated NMS factors match their scaled values at the same level of approximation. This demonstrates that it may not be appropriate to use the scaled NMS factors in high-precision determinations of nuclear charge radii by combining IS measurements with the theoretical IS factors.

\section{Global fit to optical measurements}

When combined with experimental data, our calculated IS factors enable to both extract $\delta r^2$, and test the reliability of the methods used previously to calculate IS factors in Zn~I. 
As the experimental data is of limited accuracy compared with the calculations, it is beneficial to increase the precision by combining IS measurements of different transitions in a global fit.

ISs were measured for the $D_2$ line in Zn II using a trapped sample and a laser with a wavelength of $202\,$nm~\cite{2003-Trap}.
The measured isotope pairs were: (64,66) (66,68) and (68,70).
They were also measured in a long chain of short-lived isotopes for the $481\,$nm, $4p$ $^3P_2 \rightarrow$ $5s$ $^3 S_1 $ transition at ISOLDE~\cite{2019-ISOLDE}.
The $481\,$nm results were given relative to $A=68$.
We are thus interested in projecting our calculated $202\,$nm line factors to the $481\,$nm line in order to extract $\delta r^2$ for the short-lived isotopes. To exploit the high accuracy with which we calculated $K_{202}$ and $F_{202}$, we also consider the precisely measured (as compared with the magnitude of the FS) ISs of the intercombination line at $308\,$nm~\cite{1997-308}, for which the pairs (64,66) (66,68), (68,70) and (66,67) were measured.
To be precise, we use $\delta \nu_{308}^{A,A'}$ as a bridge between $\delta \nu_{202}^{A,A'}$ for which we have accurate IS factors, and $\delta \nu_{481}^{68,A'}$ which were measured in a long chain of isotopes.

The reduced IS's of the three transitions, as well as their covariance matrices, are used as input in a fit based on Eq.~(\ref{eq:lin}). They are given in Table \ref{tab:inputs}. We assume that the raw measurements are independent and normally distributed. To account for correlations stemming from the fact that the ISs of the $481\,n$m line are given for different pairs, we introduce a mixing matrix $\mathbf{M}$ as defined in Table \ref{tab:inputs}. The matrix transforms the measurement pairs from those measured at ISOLDE to those measured for the other two lines. It thus introduces off-diagonal elements in the covariance matrix, as defined in Table~\ref{tab:inputs}. To define the fitting model, we apply Eq.~(\ref{eq:lin}) for two transition pairs to obtain the coupled linear equations
\begin{eqnarray}\label{eq.:fit}
&& \hat{y} = K'_{yx} + F_{yx} \hat{x}   \nonumber \\
\text{and} \nonumber \\
&& \hat{z} = K _{zx} + F_{zx} \hat{x},
\end{eqnarray}
where the hatted vectors indicate the adjusted positions of the values.
%
\begin{table}[tbp]
\caption{
Observed data and definition of input to global fit. 
The data are arranged as vectors with the isotope pairs indicated at the top. 
Here, $\mu$ is the corresponding inverse reduced mass vector; x, y, and z are the observed reduced ISs of the 308\,nm 481\,nm and 202\,nm lines, respectively given in Refs. \cite{1997-308}, \cite{2019-ISOLDE} and \cite{2003-Trap}; and  $\Omega_x$, $\Omega_y$ and $\Omega_z$ are the corresponding covariance matrices. The matrix $\mathbf{M}$ rearranges the reduced ISs measured for the $481\,$nm line from the (64,68) (66,68) (68,70) (67,68) isotope pairs indicated on top. It introduces off-diagonal elements in $\Omega_y$.
}

\begin{tabular}{l r r r r r l c}
\hline
\hline
  & Pair: & (64,66) & (66,68) & (68,70) & (66,67) & &\\\\
$\mu$   &  $-[$&  $4.7404$ & $4.4658$ & $4.2138$ & $2.2700$ & $]^T$  & $10^{-4}\,$u$^{-1}$\\
x   &  $[$&  $1452.6$ &$1515.1$ &$1416.8$ &$1749.3$ & $]^T$        & GHz u  \\
y   & $\mathbf{M}[$&  $153.4$ & $142.4$ & $164.9$ & $188.5$ & $]^T$        & GHz u    \\
z   &  $[$&  $1426.1$& $1500.3$& $1347.9$ & & $]^T$  & GHz u \\
$\Omega_x$ &  $[\mathrm{diag}($&$2.5$ &$3.1$& $4.3$ &$8.8$&$)]^2$ & [GHz u]$^2$\\
$\Omega_y$ & $\mathbf{M}[\mathrm{diag}($&$1.2$& $3.4$& $2.1$& $9.6$&$)]^2\mathbf{M}^T$ & [GHz u]$^2$\\
$\Omega_z$ &  $[\mathrm{diag}($&$12.7$& $8.9$ & $23.7$& &$ )]^2$  & [GHz u]$^2$
\end{tabular}
\begin{ruledtabular}
\begin{tabular}{l c c}

 & $ \mathbf{M}  = \left( \begin{array}{cccc} 
 1+\frac{\mu_2}{\mu_1}  & -\frac{\mu_2}{\mu_1} & 0 & 0 \\
 0 & 1 & 0 & 0 \\ 
 0 & 0 & 1 & 0 \\ 
 0 & \frac{\mu_2}{\mu_2-\mu_4}  & 0 & -\frac{\mu_4}{\mu_2-\mu_4} 
 \end{array}\right)$ \\
 
\end{tabular}
\end{ruledtabular}
\label{tab:inputs}
\end{table}
Note that we have added an apostrophe to $K_{yx}$ to denote that the systematic uncertainties in the beam energy of ISOLDE's measurement are absorbed in it. This is well-justified, as described in Ref.~\cite{2019-ZnThesis}.
The residuals associated with Eq. (\ref{eq.:fit}) are
\begin{eqnarray}\label{eq:res}
&& r_x = x - \hat{x}, \nonumber \\
&& r_y = y - \hat{y}  \nonumber \\
\text{and} \nonumber \\
&& r_z = z - \hat{z}_{1:3} , 
\end{eqnarray}
where we note that $r_z$ is taken only for the three measured pairs. $\delta \nu_{202}^{66,67}$ has not been measured.
To take the covariance matrices into account as generalized weights, the fit is accomplished by minimizing the squared Mahalanobis distance
\begin{eqnarray}\label{eq:Chi2}
    S(\hat{x},F,K)
    = r_x^T \Omega_x^{-1} r_x+r_y^T \Omega_y^{-1} r_y+ r_z^T \Omega_z^{-1} r_z.
\end{eqnarray}
Note that the covariance matrix $\Omega_y$ is not diagonal as we needed to match the measured pairs in the $481\,$nm line to those measured for the other two lines. Had it been diagonal, we could perform a standard York regression \cite{2004-York}. We are interested in finding the values of the four adjusted reduced ISs in $\hat{x}$, two fitted slopes $F_{yz}$ and $F_{zx}$, and two fitted intercepts $K'_{yx}$ and $K_{zx}$ that minimize the test statistic given in Eq.~(\ref{eq:Chi2}); a total of $8$ parameters.
The resulting values of these parameters are given in Table \ref{tab:outputs}. From them, the most probable values of the adjusted reduced ISs $\hat{y}$ and $\hat{z}$ are calculated using Eq.~(\ref{eq:lin}), and are also given in Table \ref{tab:outputs}.

\begin{figure}[tbb]
\centering
\includegraphics[trim={135 242 120 242},clip, width=1\columnwidth]{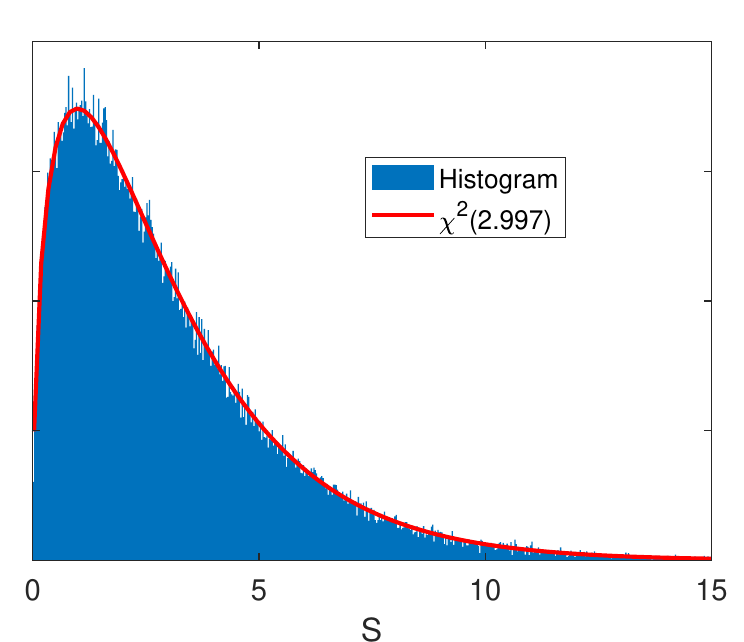}
\caption{\label{fig:Chi2}
Distribution of the squared Mahalanobis distances returned for each Monte-Carlo (MC) iteration by minimizing Eq.~(\ref{eq:Chi2}), which is fitted with a $\chi^2$ distribution.
}
\end{figure}

To estimate distributions of the adjusted and fitted parameters, we use a Monte-Carlo (MC) model of repeated measurements. For each iteration, $11$ reduced ISs were generated from a normal distribution centered at the most probable adjusted value, with a standard deviation taken as the original reported error. A function corresponding to Eq.~(\ref{eq:Chi2}) was constructed and minimized, and the resulting parameters were recorded.
%
%
A histogram of the resulting test statistic $S$ from $10^5$ iterations is shown in Fig.~\ref{fig:Chi2}. It is found to be well-described by a $\chi^2$ distribution with mean $2.997$, corresponding to our $3$ degrees of freedom well.

The outputs of the MC procedure are the posterior distributions of the adjusted reduced ISs $\hat{x}$, and the fitted slopes and intercepts.
Applying Eq.~(\ref{eq:lin}) in each iteration gives the posterior distributions of the adjusted reduced ISs $\hat{y}$ and $\hat{z}$. All the random-variables mentioned above are found to be normally distributed. Their most probable values and standard deviations are given in Table \ref{tab:outputs}.
Comparing the prior and posterior most probable values of the reduced ISs, we see small shifts as well as reduced uncertainties. For example, the reduced IS of the (66,68) mass pair for the $481\,$nm line was adjusted by the fit from $y_2=142.4(3.4)\,$GHz\,u to $\hat{y}_2=146.5(1.7)\,$GHz\,u. To see how the measured value was adjusted by the fit, we multiply by the reduced mass difference $\mu_2=-4.4658\times10^{-4}\,$u$^{-1}$, so that $\delta \nu^{66,68}_{481}=-63.6(1.5)\,$MHz shifted to $-65.5(0.8)\,$MHz; a deviation by $1.2$ reported standard error, and a reduction of the error by a factor of two.


\begin{table}[!tbp]
\caption{
Main outputs of the global fit.
The most probable values of the posterior distributions are given, with the errors indicating the standard deviations. When a distribution is asymmetric, we indicate the one-sided standard deviation.
Note that $\hat{y}$ does not include a systematic uncertainty, so we denote the relevant intercepts with a prime.}
\begin{ruledtabular}
\begin{tabular}{l r r r r }

          & (64,66)        & (66,68)         & (68,70)        & (66,67) \\
$\hat{x}$ & $1452.8(2.5)$& $1515.1(3.0)$ &$1416.4(4.1)$ &$1749.0(8.7)$\\
$\hat{y}$ & $158.5(1.2)$ & $146.5(1.7)$  &$165.5(1.8)$  &$101.4(8.2)$\\
$\hat{z}$ & $1417(10)$   & $1502(9)$     &$1367(17)$    & $1821(57)$\\

\end{tabular}
\begin{tabular}{l l  }
$F _{yx}=-0.193(29)$ & $K'_{yx}= ~~ 438(43)$ GHz\,u \\
$F_{zx}= ~~ 1.36(21)$ & $K_{zx}=-556(313)$ GHz\,u \\
$F_{yz}=-0.136^{+23}_{-35}$ & $K'_{yz}=~~351^{+54}_{-35}$ GHz\,u 
\end{tabular}
\end{ruledtabular}
\label{tab:outputs}
\end{table}

We also calculate the distributions of the intercept and slope for the unfitted transition pair by solving 
\begin{eqnarray}\label{eq:unfit}
&& F_{yz} = F_{yx}/F_{zx} \nonumber \\
\text{and} && \nonumber \\
&& K'_{yz}=  K'_{yx}-F_{yz} K_{zx}, 
\end{eqnarray}
in each MC iteration, in accordance with Eq.~(\ref{eq:lin}).
The resulting distributions are plotted in Fig.~\ref{fig:KF}. They are found to be highly asymmetric. Nevertheless, they fit reasonably well to one-sided Gaussian distributions. We, thus, report the most probable values and single-sided standard deviations in Table \ref{tab:outputs}. To show the importance of including $\delta \nu_{308}^{A,A'}$ in the fit, we perform an analogous MC fit without it by taking as inputs $y_{1:3}$ and $z$ and their covariance matrices. The resulting distributions of the intercept and slope are also plotted in Fig.~\ref{fig:KF}. They are found to be even more asymmetric, with tails that are not well-described by one-sided normal distributions. The one-sided uncertainty is also increased by as much as a factor of two. We conclude that the inclusion of $\delta \nu_{308}^{A,A'}$ in the fit is useful for reducing the uncertainties, obtaining well-behaved distributions, and for estimating the unmeasured IS $\delta \nu_{202}^{66,67}$.

The self-consistent fitting procedure described above is both straightforward and general; it may be employed to ISs measured for any number of transitions, with each transition measured for a different number of pairs and for different choices of pairs of isotopes.

\section{Testing calculations in Zn~I}

\begin{figure}[tb]
\centering
\includegraphics[trim={0 200 43 205},clip, width=0.87\columnwidth]{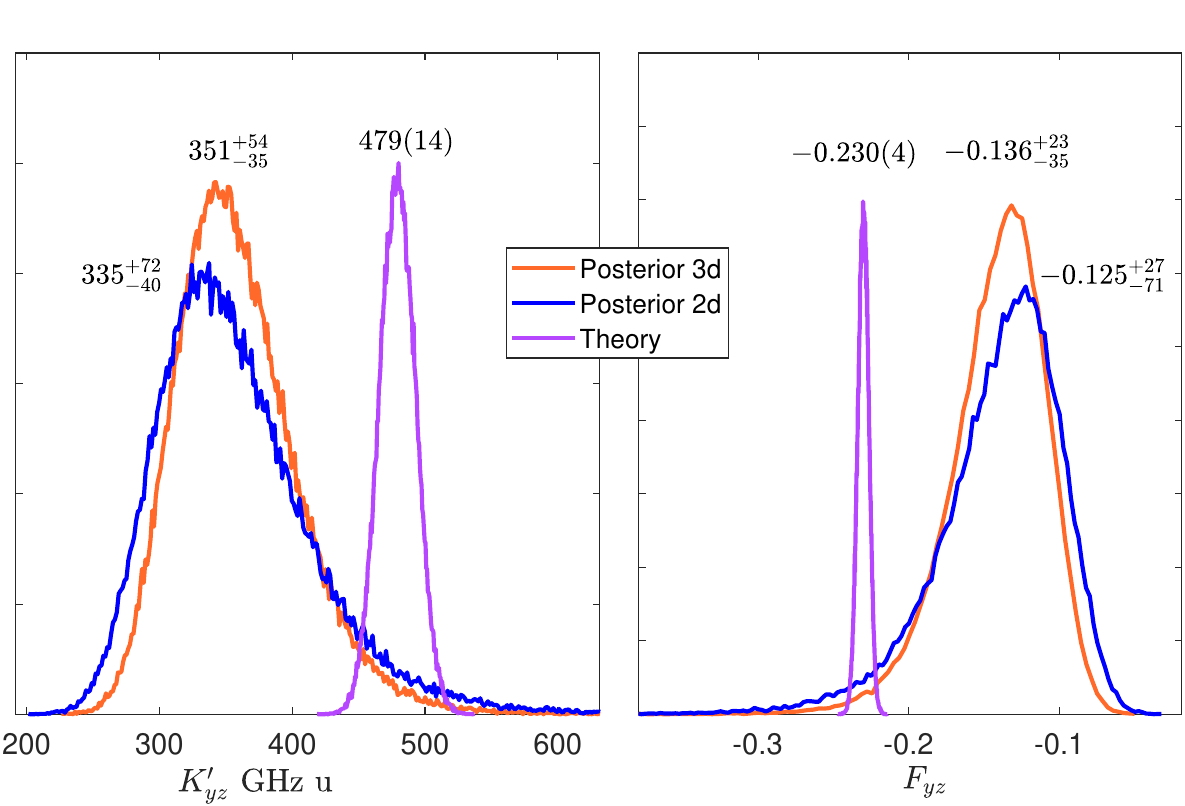}
\caption{\label{fig:KF}
Posterior distributions of the slope and intercept connecting the reduced ISs of the $202\,$nm and $481\,$nm lines.
They are estimated by applying Eq.~(\ref{eq:unfit}) to the fitted parameters in each MC iteration. 
The graphs denoted ``3d" correspond to the full global fit, while those denoted ``2d" correspond to a fit which does not include the $308\,$nm results.
The predicted values are obtained by combining $F_{202}$ and $K_{202}$ from this work with $F_{481}$ and $K_{481}$ from Ref. \cite{2017-MCDHF}. A systematic uncertainty of $7.3\,$GHz is added to the theoretical curve of $K'_{yz}$.
}
\end{figure}

We compare the fitting results with their theoretical predictions, given by a combination of $K_{202}$ and $F_{202}$ calculated in this work, and $K_{481}$ and $F_{481}$ calculated via the multiconfiguration Dirac-Hartree-Fock (MCDHF) method from Ref.~\cite{2017-MCDHF}. All are assumed to be independent and normally distributed. The resulting distributions are plotted in Fig.~\ref{fig:KF}. To account for the systematic uncertainty in the experiment, we add an uncertainty of $7.3\,$GHz~u in quadrature to the theoretical prediction of $K'_{zy}$.
We see a difference between the most probable values of experiment and theory. It is on the level of $2.4$ combined standard deviations for the intercept, and $2.7$ combined standard deviations for the slope.
As we are not anticipating major changes in the calculated IS factors of Zn II, and we have taken a conservative estimate of their uncertainties, we infer that there is a deviation between experiment, and the calculated factors of the $481\,$nm line.
This could be due to omission of some missing physical effects in the MCDHF calculation.
It is worth mentioning here that the MCDHF method cannot account for the core-polarization effects as rigorously as the RCC method~\cite{Kaldor,bijaya3}.

To further test this calculation, we project the calculated factors $K_{202}$ and $F_{202}$ to the measured transitions in Zn~II.
This is done by considering distributions of 
\begin{eqnarray*}
 && F_{308} =~ F_{202}/F_{zx} \\
 && F_{481} =~ F_{202} F_{yz} \\
 && K_{308} =~ K_{zx}+F_{zx}K_{202} \\
 \text{and} && \nonumber \\
 && K'_{481} =~ K'_{yz}+F_{yz}K_{202}.
\end{eqnarray*}

The corresponding most probable values, with one-sided standard deviations, are given in Table \ref{tab:ISproj}. Their uncertainties are dominated by the statistical experimental uncertainties in the ISs of the $481\,$nm and $202\,$nm lines.
For the IS factors of the 308\,nm line, our projected values agree well, within the rather large experimental uncertainty bounds, with the calculated values of Ref.~\cite{2017-MCDHF}. 
However, when using the fit results, namely $K'_{yx}$ and $F_{yx}$ from Table~\ref{tab:outputs}, to project the calculated values of the $481\,$nm line factors~\cite{2017-MCDHF}, we find deviations of $2-3$ combined standard deviations. In other words, irrespective of the calculations reported in this work, the IS factors for the two transitions given in~\cite{2017-MCDHF} are incompatible with each other. This fact has been pointed out in Ref.~\cite{2017-MCDHF}. Similarly, for the 481 line, our projected IS factors for the 202 nm line agree with the projected factors of the 308 nm line from Ref.~\cite{2017-MCDHF}, and disagree with those from the previously reported {\it ab initio} calculations. For $K_{481}$, the deviation is $2.6$ combined standard deviation. A similar deviation is found for $F_{481}$, whereas for the MS factor, in which strong cancellation occurs, such deviation may not be surprising. For the FS factor, for which different calculations usually agree to within a few percent, such deviation was unexpected.

\begin{table}[tbp]
\caption{
Comparison of the projected and calculated IS factors.
The most probable values of the posterior distribution are given, with the errors indicating the standard deviations. When a distribution is asymmetric, we indicate the one-sided standard deviation.
A global systematic uncertainty is added to the factors of the $481\,$nm transition which rely on the IS measurements. It is indicated with square brackets.
} 
\begin{ruledtabular}
\begin{tabular}{l l l l  }
        & $F_i\,$MHz/fm$^2$ & $K_i\,$GHz u & Method\\
        \hline \\
 308 nm &  $-1107(173)$ & $1902^{+78}_{-51}$ & AR-CCSDT(202)+fit\\
        &  $-1135(4)$   & $1955(19)$ & MCDHF(308)~\cite{2017-MCDHF}\\
         & $-1796(271)$ & $2159^{+136}_{-80}$ & MCDHF(481)+fit\\
 \\
 481 nm & ~~ $204^{+53}_{-34}$ & $75^{+14}_{-21}[7.3]$ & AR-CCSDT(202)+fit\\
        & ~~ $219(33)$ & $64(14)[7.3]$ & MCDHF(308)+fit \\
        & ~~ $346(3)$  & $14(7)$ & MCDHF(481)~\cite{2017-MCDHF}\\
        & ~~ $346(35)$ & $49(17)[7.3]$ & MCDHF(481)+MuX~\cite{2019-ISOLDE}
\end{tabular}
\end{ruledtabular}
\label{tab:ISproj}
\end{table}

\section{$\delta r^2$ of muonic and electronic Zn}

\subsection{Stable Nuclei}

In each MC iteration, the differential radii of the stable isotope pairs may be deduced directly from our calculated factors via
\begin{equation}
\delta r^2 =(\hat{z}-K_{202})\mu/F_{202}.
\label{IS}
\end{equation}
The resulting distributions are found to be normally distributed. Their most probable values and standard deviation are given in Table~\ref{tab:rad}, with uncertainties of order $0.008\,$fm$^2$. They arise, except for the pair $(66,67)$, from the error of $K_{202}$, in turn stemming from unaccounted-for electron correlation contributions to $K_{202}^\mathrm{SMS}$.

It is customary to test atomic many-body calculations by comparing the optically-determined radii to those extracted from muonic atoms, whose energy levels are highly sensitive to nuclear effects. In muonic Zn, cascade X-ray energies were measured by Shera~\textit{et. al.} in 1976~\cite{1976-Shera}, and interpreted as Barret equivalent radii $R^k_\alpha$, which in spherical nuclei are considered nearly free from nuclear model dependency~\cite{1969-Barret}.
To extract RMS radii from $R^k_\alpha$, knowledge of the nuclear charge distribution is needed. In even-even, stable, isotopes Zn, these distributions were determined by Wohlfart~\textit{et. al.} via elastic electron scattering, fitted with a Fourier-Bessel series~\cite{1980-MuE}. 
The same group performed a combined analysis of the electron scattering and muonic X-ray data, and extracted $\delta r^2$. They included a large uncertainty consistently, stemming from the finite accuracy of the scattering experiment. They did not include, however, any error associated with the nuclear polarization. These results are quoted in Table~\ref{tab:rad}.

\begin{table}[tbp]
\caption{
Differential radii $\delta r^2$ for stable Zn nuclei in fm$^2$. 
In this work, they are obtained from the global fit in combination with the calculated IS parameters.
We also give our recommended values for the radii extracted via muonic atoms X-ray spectroscopy, including large uncertainties whose estimations are described in the main text.
} 
\begin{ruledtabular}
\begin{tabular}{l r r r r}

\multicolumn{1}{r}{Pair:}  &
\multicolumn{1}{c}{(64,66)} &
\multicolumn{1}{c}{(66,68)}  &
\multicolumn{1}{c}{(68,70)} &
\multicolumn{1}{c}{(66,67)} \\ 
\hline \\ 
Optical          & $0.191(08)$ & $0.154(07)$ &$0.183(08)$ & $0.031(9)$\\
Muonic & $0.162(28)$ & $0.131(30)$ &$0.149(25)$\\
\cite{1980-MuE}    & $0.177(24)$ & $0.114(24)$ & $0.138(20)$ \\
\cite{1982-Deform} & $0.158(03)$ & $0.129(03)$ & $0.148(08)$ \\
\cite{2004-FH}     & $0.152(06)$ & $0.149(06)$ & $0.160(11)$ 
\end{tabular}
\end{ruledtabular}
\label{tab:rad}
\end{table}

Foot \textit{et. al.}~\cite{1982-Deform} plotted the aforementioned $ \delta r^2 $ values against the optical IS data using Eq.~(\ref{eq:IS}) and found an inconsistency between the two. For this reason, they opted to use the Coulomb excitation data~\cite{1962-SM} and a simple model for nuclear deformation to interpret the muonic data. The resulting $\delta r^2$ were found to be more consistent with the optical data. They are also given in Table~\ref{tab:rad} with their originally quoted uncertainties, which are only statistical. 
We note that application of Eq.~(\ref{eq:IS}) can only test optical vs. muonic data up to a linear transformation, while a direct comparison of $\delta r^2$ may be more discerning.

In their seminal review~\cite{2004-FH}, Fricke and Heilig performed a combined analysis of the same electron scattering and muonic X-ray data used by Wohlfart~\textit{et. al.}. Their results are given in Table~\ref{tab:rad}. Here, only statistical and nuclear polarization uncertainties were included. 
Considering that a similar analysis procedure was used, the results of Refs.~\cite{2004-FH} and \cite{1980-MuE} are surprisingly dissimilar.
This indicates that experimental errors in the electron scattering data propagate through the combined analysis into the determined $ \delta r^2$ and must be taken into account. 

Considering the above, our recommended values for $\delta r^2$ extracted from the muonic data are given in Table~\ref{tab:rad}. This is given as the average of three determinations, with all uncertainties added in quadrature, including an allowance for the variation in the results between different interpretations. We presume this is mainly due to the nuclear model dependency.
The resulting uncertainties of order $0.03\,$fm$^2$ may seem excessive, especially as compared with the aggressive ones given in the latest compilation~\cite{2013-AM}. However, their magnitudes may be understood in the following way. Both the Barret and the absolute radii quoted in three analyses~\cite{1980-MuE, 1982-Deform, 2004-FH} differ by a few per mill, indicating the uncertainty in the choice and application of the nuclear model. In simple spherical nuclei close to shell closures, the nuclear model uncertainty would largely cancel for the differential radii. However, stable Zn nuclei are rather deformed, with the deformation parameters varying between the different isotopes. We may thus expect uncertainties in the differential radii to be of the order of $(r\approx4~\mathrm{fm})^2 \times0.1\%\approx0.02$~fm$^2$, in line with those given in Table~\ref{tab:rad}, which also include experimental and nuclear polarization errors.

Having established that the extraction of $ \delta r^2 $ from muonic atoms is subject to large uncertainties stemming from the nuclear model, it is natural to ask how large is this effect for the optical $ \delta r^2 $ extracted from Eq.~(\ref{eq:IS}) in this work. The main isotope-dependant correction to Eq.~(\ref{eq:IS}) is considered to be proportional to the changes in the fourth moments of the charge distribution, taking the form $S_4 \delta r^{(4)}$, with $S_4=-3.3\times10^{-4}~$fm$^{-2}$ for the ground state in Zn given in Ref.~\cite{1987-Blund}.
For the (68,70) isotope pair, we can calculate $\delta r^{(4)}=3.4\pm3.2~$fm$^4$ from the differential fourth moments given, with a large uncertainty, in Ref.~\cite{1980-MuE}, resulting in a correction whose magnitude is $-1\pm1~$fm$^2$. This correction is completely negligible given the uncertainty quoted in Table~\ref{tab:rad}.
This demonstrates that optical determinations of $\delta r^2$ depend much less on our assumptions concerning minute isotopic variations in the nuclear shape.
Comparing $ \delta r^2 $ deduced from optical measurements and muonic atoms, we see an agreement within one combined error, with the precision of the optical measurement a factor $3-4$ times higher.
Thus, it is the optical determination that tests the muonic one and not the other way around. 

\subsection{Short-lived Nuclei}

\begin{figure*}[!tbb]
\includegraphics[trim={0 212 0 238},clip,width=1.97\columnwidth]{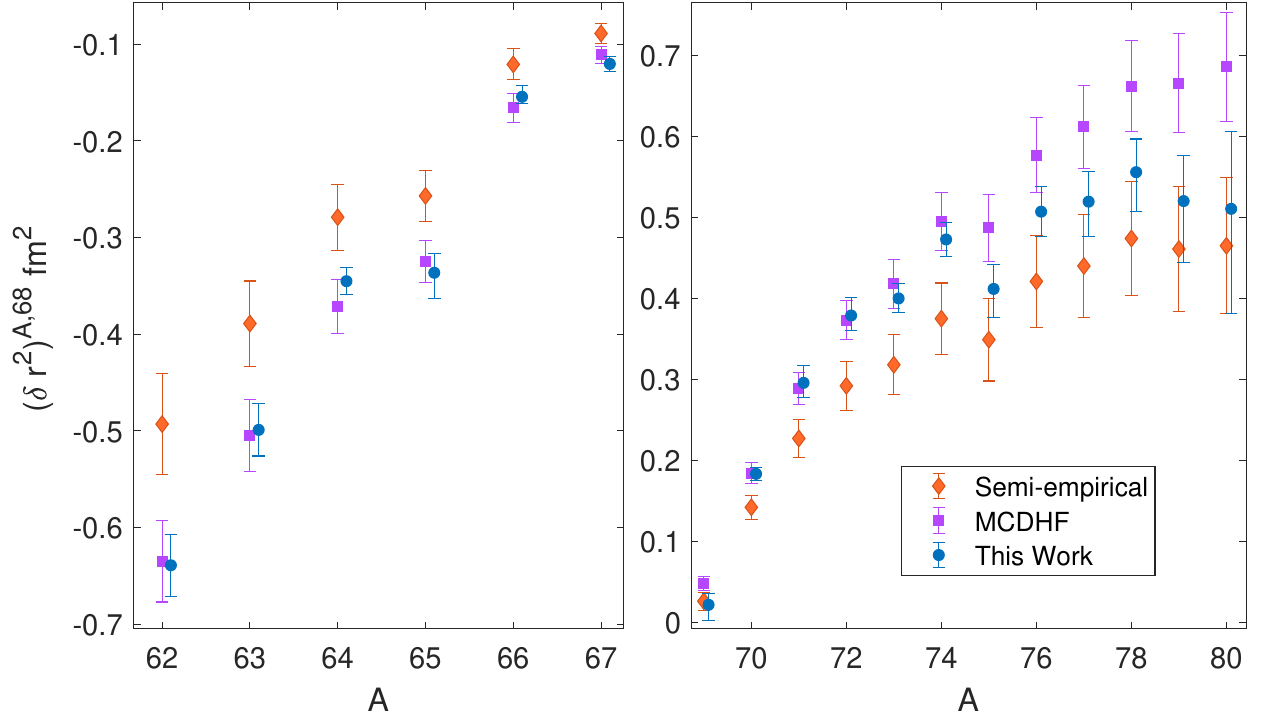}
\caption{\label{fig:rad}
Differential radii in the Zn isotopic chain, given in Table~\ref{tab:Rad2}. Here, the left panel indicates proton rich nuclei while the right panel shows the neutron rich ones.}
\end{figure*}
Both the muonic $ \delta r^2 $ given in Table~\ref{tab:rad}, and the calculated $K_{481}$ and $F_{481}$ given in Table~\ref{tab:factors} show deviations and inconsistencies.
These in turn have propagated to the extraction of $\delta r^2 $ for the short-lived Zn nuclei from the measurements of Ref.~\cite{2019-ISOLDE}.

The $\delta r^2$ deduced when using the calculated IS factors for the 481\,nm line are given in Table~\ref{tab:Rad2} and plotted in Fig.~\ref{fig:rad}. Their main uncertainty contribution comes from the systematic uncertainty related to the calibration of the extraction voltage. These $\delta r^2$ were regarded unrealistic when considering the behavior of neighboring chains. In light of this, the authors of Ref.~\cite{2019-ISOLDE} chose a semi-empirical strategy. First, they adopted the calculated $F_{481}$ with an increased uncertainty of $10\%$, then they used independent $ \delta r^2 $ of stable even isotopes from Ref.~\cite{2004-FH} to deduce $K'_{481}=49(17)$. Comparing these factors with the rest in Table~\ref{tab:factors}, we see that $K'_{481}$ is indeed within one standard deviation of our recommended value. The calibration strategy was thus successful in mitigating the deviation in $K_{481}$. However, the deviation in $F_{481}$ remains, causing significant shifts in the determined $\delta r^2$, also given in Table~\ref{tab:Rad2} and plotted in Fig.~\ref{fig:rad}.

Our recommended values for $\delta r^2$ are based on the distributions of the projected IS factors $F_{481}$ and $K'_{481}$ whose most probable values are given in Table~\ref{tab:factors}. 
They are given in Table~\ref{tab:Rad2} and plotted in Fig.~\ref{fig:rad}.
As can be seen from the table, their uncertainties are largely dominated by those of the projected IS factors, in turn stemming from statistical uncertainties in the stable isotopes IS measurements of the $202~$nm and $481~$nm lines. 
Our recommended $\delta r^2$ for the chain agree with those from the direct theoretical determinations close to stability ($62\leq A \leq 74$), and disagree for the most neutron rich isotopes. The situation is opposite for the semi-empirical extraction.
Adopting either one of the other results would result in deviations from our recommended values by as much as 3 times their reported errors.

\begin{table}[!htbp]
\centering
\caption{
Differential radii $ (\delta r^2 )^{A,68}$ in fm$^2$ for the Zn isotopic chain extracted from the IS measurements of Ref.~\cite{2019-ISOLDE} using IS factors from different works. They are also shown in Fig.~\ref{fig:rad}.}
\begin{ruledtabular}
\begin{tabular}{crrr}

A & This Work & Semi-empirical~\cite{2019-ISOLDE} & MCDHF~\cite{2017-MCDHF}\\
\hline \\
62 & $-0.639(32)$          & $-0.493(52)$ & $-0.635(42)$\\
63 & $-0.499(27)$          & $-0.389(44)$ & $-0.505(37)$\\
64 & $-0.345(14)$          & $-0.279(34)$ & $-0.371(28)$\\
65 & $-0.337^{+20}_{-27}~$ & $-0.257(26)$ & $-0.325(22)$\\
66 & $-0.154(07)$          & $-0.121(16)$ & $-0.166(15)$\\
67 & $-0.121(08)$          & $-0.089(10)$ & $-0.111(09)$\\
69 & $ 0.022^{+14}_{-20}~$ & $ 0.026(11)$ & $ 0.048(09)$\\
70 & $ 0.183(08)$          & $ 0.142(15)$ & $ 0.184(13)$\\
71 & $ 0.296^{+22}_{-18}~$ & $ 0.227(24)$ & $ 0.289(20)$\\
72 & $ 0.379^{+22}_{-19}~$ & $ 0.292(30)$ & $ 0.373(24)$\\
73 & $ 0.400(18)$          & $ 0.318(37)$ & $ 0.418(30)$\\
74 & $ 0.473(21)$          & $ 0.375(44)$ & $ 0.495(36)$\\
75 & $ 0.412^{+30}_{-35}~$ & $ 0.349(51)$ & $ 0.487(41)$\\
76 & $ 0.507(31)$          & $ 0.421(57)$ & $ 0.577(46)$\\
77 & $ 0.520^{+37}_{-43}~$ & $ 0.440(64)$ & $ 0.612(51)$\\
78 & $ 0.556^{+41}_{-49}~$ & $ 0.474(70)$ & $ 0.662(56)$\\
79 & $ 0.520^{+56}_{-76}~$ & $ 0.461(77)$ & $ 0.666(61)$\\
80 & $ 0.511^{+96}_{-129}$ & $ 0.465(84)$ & $ 0.686(67)$
\end{tabular}
\end{ruledtabular}
\label{tab:Rad2}
\end{table}

\section{Summary and outlook}

We performed state-of-the-art calculations of energies and isotope shift constants of the first three low-lying states of Zn~II. A remarkable agreement of the calculated energies with the experimental values is found; at the level of few hundred parts per million. For the FS and SMS factors, our calculations agree, and are more precise than, another calculation carried out using the combined configuration interaction method and many-body perturbation theory. The first-principle studies of the NMS constants in Zn are carried out and compared with the values obtained using the scaling-law. Significant deviations are observed, because of which we have used the calculated values for the nuclear radii analysis.

A global fit approach is adopted with the measurements of isotope shifts with our calculated isotope 
shift factors of Zn II to infer isotope shift constants of Zn I and differential nuclear charge radii of Zn isotope chain. We compared the projected isotope shift factors with those recently calculated using the multiconfiguration Dirac-Hartree-Fock method. It showed an agreement for the factors of the $308\,$nm line and a significant deviation for those of the $481\,$nm line of Zn I. This explains why the radii extracted for the Zn isotope chain using the factors of the $481\,$nm line disagreed with those of neighboring chains. We also find deviations between the radii extracted from our all-optical determination and those resulting from measurements in muonic atoms, which were hitherto believed to be quite reliable. The deviations are traced to the deformed nature of stable Zn nuclei, making the muonic determination nuclear-model dependent.
We infer that for the purpose of extracting differential radii, the combination of measurements and state-of-the-art calculations in simple transitions, projected to the ones measured online, may be the most appropriate choice.
 
%
This work may motivate others to carry out a modern analysis of the muonic Zn energy data using state-of-the-art non-perturbative QED calculations, combined with inputs from modern low-energy nuclear theory.
We also intend to extend our study to Cu I and Ga I for elucidating the behavior of differential radii in this region of the nuclear chart.





\begin{acknowledgments}
We thank  F.O.A.P. Gustafsson and G. Ron for useful comments, and D.G. Hitlin for illuminating discussions.
B.K.S. acknowledges the use of ParamVikram-1000 HPC facility at PRL Ahmedabad, for carrying out the computations of the results reported in this paper. 
B.O. is thankful for the support of the Council for Higher Education Program for Hiring Outstanding Faculty Members in Quantum Science and Technology.
\end{acknowledgments}

\bibliography{references}

\begin{thebibliography}{38}%
\makeatletter
\providecommand \@ifxundefined [1]{%
 \@ifx{#1\undefined}
}%
\providecommand \@ifnum [1]{%
 \ifnum #1\expandafter \@firstoftwo
 \else \expandafter \@secondoftwo
 \fi
}%
\providecommand \@ifx [1]{%
 \ifx #1\expandafter \@firstoftwo
 \else \expandafter \@secondoftwo
 \fi
}%
\providecommand \natexlab [1]{#1}%
\providecommand \enquote  [1]{``#1''}%
\providecommand \bibnamefont  [1]{#1}%
\providecommand \bibfnamefont [1]{#1}%
\providecommand \citenamefont [1]{#1}%
\providecommand \href@noop [0]{\@secondoftwo}%
\providecommand \href [0]{\begingroup \@sanitize@url \@href}%
\providecommand \@href[1]{\@@startlink{#1}\@@href}%
\providecommand \@@href[1]{\endgroup#1\@@endlink}%
\providecommand \@sanitize@url [0]{\catcode `\\12\catcode `\$12\catcode
  `\&12\catcode `\#12\catcode `\^12\catcode `\_12\catcode `\%12\relax}%
\providecommand \@@startlink[1]{}%
\providecommand \@@endlink[0]{}%
\providecommand \url  [0]{\begingroup\@sanitize@url \@url }%
\providecommand \@url [1]{\endgroup\@href {#1}{\urlprefix }}%
\providecommand \urlprefix  [0]{URL }%
\providecommand \Eprint [0]{\href }%
\providecommand \doibase [0]{http://dx.doi.org/}%
\providecommand \selectlanguage [0]{\@gobble}%
\providecommand \bibinfo  [0]{\@secondoftwo}%
\providecommand \bibfield  [0]{\@secondoftwo}%
\providecommand \translation [1]{[#1]}%
\providecommand \BibitemOpen [0]{}%
\providecommand \bibitemStop [0]{}%
\providecommand \bibitemNoStop [0]{.\EOS\space}%
\providecommand \EOS [0]{\spacefactor3000\relax}%
\providecommand \BibitemShut  [1]{\csname bibitem#1\endcsname}%
\let\auto@bib@innerbib\@empty
\bibitem [{\citenamefont {Dilling}\ \emph {et~al.}(2018)\citenamefont
  {Dilling}, \citenamefont {Blaum}, \citenamefont {Brodeur},\ and\
  \citenamefont {Eliseev}}]{2018-Traps}%
  \BibitemOpen
  \bibfield  {author} {\bibinfo {author} {\bibfnamefont {Jens}\ \bibnamefont
  {Dilling}}, \bibinfo {author} {\bibfnamefont {Klaus}\ \bibnamefont {Blaum}},
  \bibinfo {author} {\bibfnamefont {Maxime}\ \bibnamefont {Brodeur}}, \ and\
  \bibinfo {author} {\bibfnamefont {Sergey}\ \bibnamefont {Eliseev}},\
  }\bibfield  {title} {\enquote {\bibinfo {title} {Penning-trap mass
  measurements in atomic and nuclear physics},}\ }\href {\doibase
  10.1146/annurev-nucl-102711-094939} {\bibfield  {journal} {\bibinfo
  {journal} {Annual Review of Nuclear and Particle Science}\ }\textbf {\bibinfo
  {volume} {68}},\ \bibinfo {pages} {45--74} (\bibinfo {year} {2018})},\
  \Eprint
  {http://arxiv.org/abs/https://doi.org/10.1146/annurev-nucl-102711-094939}
  {https://doi.org/10.1146/annurev-nucl-102711-094939} \BibitemShut {NoStop}%
\bibitem [{\citenamefont {Yang}\ \emph {et~al.}(2023)\citenamefont {Yang},
  \citenamefont {Wang}, \citenamefont {Wilkins},\ and\ \citenamefont
  {Ruiz}}]{2023-Review}%
  \BibitemOpen
  \bibfield  {author} {\bibinfo {author} {\bibfnamefont {X.F.}\ \bibnamefont
  {Yang}}, \bibinfo {author} {\bibfnamefont {S.J.}\ \bibnamefont {Wang}},
  \bibinfo {author} {\bibfnamefont {S.G.}\ \bibnamefont {Wilkins}}, \ and\
  \bibinfo {author} {\bibfnamefont {R.F.~Garcia}\ \bibnamefont {Ruiz}},\
  }\bibfield  {title} {\enquote {\bibinfo {title} {Laser spectroscopy for the
  study of exotic nuclei},}\ }\href {\doibase
  https://doi.org/10.1016/j.ppnp.2022.104005} {\bibfield  {journal} {\bibinfo
  {journal} {Progress in Particle and Nuclear Physics}\ }\textbf {\bibinfo
  {volume} {129}},\ \bibinfo {pages} {104005} (\bibinfo {year}
  {2023})}\BibitemShut {NoStop}%
\bibitem [{\citenamefont {Fricke}()}]{2004-FH}%
  \BibitemOpen
  \bibfield  {author} {\bibinfo {author} {\bibfnamefont {K.}~\bibnamefont
  {Fricke}, \bibfnamefont {G.~nand~Heilig}},\ }\bibfield  {title} {\enquote
  {\bibinfo {title} {Nuclear {C}harge {R}adii {\textperiodcentered} 30-{Z}n
  {Z}inc: {D}atasheet from {L}andolt-{B{\"o}}rnstein - {G}roup {I} {E}lementary
  {P}articles, {N}uclei and {A}toms {\textperiodcentered} volume 20: ``nuclear
  charge radii''},}\ }\href {\doibase 10.1007/10856314\_32} {\
  10.1007/10856314\_32},\ \bibinfo {note} {copyright 2004 Springer-Verlag
  Berlin Heidelberg}\BibitemShut {NoStop}%
\bibitem [{\citenamefont {Delaunay}\ \emph {et~al.}(2017)\citenamefont
  {Delaunay}, \citenamefont {Frugiuele}, \citenamefont {Fuchs},\ and\
  \citenamefont {Soreq}}]{2017-Yotam}%
  \BibitemOpen
  \bibfield  {author} {\bibinfo {author} {\bibfnamefont {C\'edric}\
  \bibnamefont {Delaunay}}, \bibinfo {author} {\bibfnamefont {Claudia}\
  \bibnamefont {Frugiuele}}, \bibinfo {author} {\bibfnamefont {Elina}\
  \bibnamefont {Fuchs}}, \ and\ \bibinfo {author} {\bibfnamefont {Yotam}\
  \bibnamefont {Soreq}},\ }\bibfield  {title} {\enquote {\bibinfo {title}
  {Probing new spin-independent interactions through precision spectroscopy in
  atoms with few electrons},}\ }\href {\doibase 10.1103/PhysRevD.96.115002}
  {\bibfield  {journal} {\bibinfo  {journal} {Phys. Rev. D}\ }\textbf {\bibinfo
  {volume} {96}},\ \bibinfo {pages} {115002} (\bibinfo {year}
  {2017})}\BibitemShut {NoStop}%
\bibitem [{\citenamefont {Sailer}\ \emph {et~al.}(2022)\citenamefont {Sailer},
  \citenamefont {Debierre}, \citenamefont {Harman}, \citenamefont {Hei{\ss}e},
  \citenamefont {K{\"o}nig}, \citenamefont {Morgner}, \citenamefont {Tu},
  \citenamefont {Volotka}, \citenamefont {Keitel}, \citenamefont {Blaum} \emph
  {et~al.}}]{2022-g}%
  \BibitemOpen
  \bibfield  {author} {\bibinfo {author} {\bibfnamefont {Tim}\ \bibnamefont
  {Sailer}}, \bibinfo {author} {\bibfnamefont {Vincent}\ \bibnamefont
  {Debierre}}, \bibinfo {author} {\bibfnamefont {Zolt{\'a}n}\ \bibnamefont
  {Harman}}, \bibinfo {author} {\bibfnamefont {Fabian}\ \bibnamefont
  {Hei{\ss}e}}, \bibinfo {author} {\bibfnamefont {Charlotte}\ \bibnamefont
  {K{\"o}nig}}, \bibinfo {author} {\bibfnamefont {Jonathan}\ \bibnamefont
  {Morgner}}, \bibinfo {author} {\bibfnamefont {Bingsheng}\ \bibnamefont {Tu}},
  \bibinfo {author} {\bibfnamefont {Andrey~V}\ \bibnamefont {Volotka}},
  \bibinfo {author} {\bibfnamefont {Christoph~H}\ \bibnamefont {Keitel}},
  \bibinfo {author} {\bibfnamefont {Klaus}\ \bibnamefont {Blaum}},  \emph
  {et~al.},\ }\bibfield  {title} {\enquote {\bibinfo {title} {Measurement of
  the bound-electron g-factor difference in coupled ions},}\ }\href@noop {}
  {\bibfield  {journal} {\bibinfo  {journal} {Nature}\ }\textbf {\bibinfo
  {volume} {606}},\ \bibinfo {pages} {479--483} (\bibinfo {year}
  {2022})}\BibitemShut {NoStop}%
\bibitem [{\citenamefont {Frugiuele}\ and\ \citenamefont
  {Peset}(2022)}]{2022-Clara}%
  \BibitemOpen
  \bibfield  {author} {\bibinfo {author} {\bibfnamefont {Claudia}\ \bibnamefont
  {Frugiuele}}\ and\ \bibinfo {author} {\bibfnamefont {Clara}\ \bibnamefont
  {Peset}},\ }\bibfield  {title} {\enquote {\bibinfo {title} {Muonic vs
  electronic dark forces: a complete eft treatment for atomic spectroscopy},}\
  }\href@noop {} {\bibfield  {journal} {\bibinfo  {journal} {Journal of High
  Energy Physics}\ }\textbf {\bibinfo {volume} {2022}},\ \bibinfo {pages}
  {1--24} (\bibinfo {year} {2022})}\BibitemShut {NoStop}%
\bibitem [{\citenamefont {Palmer}(1987)}]{1987-Palmer}%
  \BibitemOpen
  \bibfield  {author} {\bibinfo {author} {\bibfnamefont {C~W~P}\ \bibnamefont
  {Palmer}},\ }\bibfield  {title} {\enquote {\bibinfo {title} {Reformulation of
  the theory of the mass shift},}\ }\href {\doibase
  10.1088/0022-3700/20/22/011} {\bibfield  {journal} {\bibinfo  {journal}
  {Journal of Physics B: Atomic and Molecular Physics}\ }\textbf {\bibinfo
  {volume} {20}},\ \bibinfo {pages} {5987} (\bibinfo {year}
  {1987})}\BibitemShut {NoStop}%
\bibitem [{\citenamefont {Cheal}\ \emph {et~al.}(2012)\citenamefont {Cheal},
  \citenamefont {Cocolios},\ and\ \citenamefont {Fritzsche}}]{2012-Co}%
  \BibitemOpen
  \bibfield  {author} {\bibinfo {author} {\bibfnamefont {B.}~\bibnamefont
  {Cheal}}, \bibinfo {author} {\bibfnamefont {T.~E.}\ \bibnamefont {Cocolios}},
  \ and\ \bibinfo {author} {\bibfnamefont {S.}~\bibnamefont {Fritzsche}},\
  }\bibfield  {title} {\enquote {\bibinfo {title} {Laser spectroscopy of
  radioactive isotopes: Role and limitations of accurate isotope-shift
  calculations},}\ }\href {\doibase 10.1103/PhysRevA.86.042501} {\bibfield
  {journal} {\bibinfo  {journal} {Phys. Rev. A}\ }\textbf {\bibinfo {volume}
  {86}},\ \bibinfo {pages} {042501} (\bibinfo {year} {2012})}\BibitemShut
  {NoStop}%
\bibitem [{\citenamefont {Campbell}\ \emph {et~al.}(1997)\citenamefont
  {Campbell}, \citenamefont {Billowes},\ and\ \citenamefont
  {Grant}}]{1997-308}%
  \BibitemOpen
  \bibfield  {author} {\bibinfo {author} {\bibfnamefont {P}~\bibnamefont
  {Campbell}}, \bibinfo {author} {\bibfnamefont {J}~\bibnamefont {Billowes}}, \
  and\ \bibinfo {author} {\bibfnamefont {I~S}\ \bibnamefont {Grant}},\
  }\bibfield  {title} {\enquote {\bibinfo {title} {The specific mass shift of
  the zinc atomic ground state},}\ }\href {\doibase
  10.1088/0953-4075/30/10/010} {\bibfield  {journal} {\bibinfo  {journal}
  {Journal of Physics B: Atomic, Molecular and Optical Physics}\ }\textbf
  {\bibinfo {volume} {30}},\ \bibinfo {pages} {2351} (\bibinfo {year}
  {1997})}\BibitemShut {NoStop}%
\bibitem [{\citenamefont {Xie}\ \emph {et~al.}(2019)\citenamefont {Xie},
  \citenamefont {Yang}, \citenamefont {Wraith}, \citenamefont {Babcock},
  \citenamefont {Bieroń}, \citenamefont {Billowes}, \citenamefont {Bissell},
  \citenamefont {Blaum}, \citenamefont {Cheal}, \citenamefont {Filippin},
  \citenamefont {Flanagan}, \citenamefont {{Garcia Ruiz}}, \citenamefont
  {Gins}, \citenamefont {Gaigalas}, \citenamefont {Godefroid}, \citenamefont
  {Gorges}, \citenamefont {Grob}, \citenamefont {Heylen}, \citenamefont
  {Jönsson}, \citenamefont {Kaufmann}, \citenamefont {Kowalska}, \citenamefont
  {Krämer}, \citenamefont {Malbrunot-Ettenauer}, \citenamefont {Neugart},
  \citenamefont {Neyens}, \citenamefont {Nörtershäuser}, \citenamefont
  {Otsuka}, \citenamefont {Papuga}, \citenamefont {Sánchez}, \citenamefont
  {Tsunoda},\ and\ \citenamefont {Yordanov}}]{2019-ISOLDE}%
  \BibitemOpen
  \bibfield  {author} {\bibinfo {author} {\bibfnamefont {L.}~\bibnamefont
  {Xie}}, \bibinfo {author} {\bibfnamefont {X.F.}\ \bibnamefont {Yang}},
  \bibinfo {author} {\bibfnamefont {C.}~\bibnamefont {Wraith}}, \bibinfo
  {author} {\bibfnamefont {C.}~\bibnamefont {Babcock}}, \bibinfo {author}
  {\bibfnamefont {J.}~\bibnamefont {Bieroń}}, \bibinfo {author} {\bibfnamefont
  {J.}~\bibnamefont {Billowes}}, \bibinfo {author} {\bibfnamefont {M.L.}\
  \bibnamefont {Bissell}}, \bibinfo {author} {\bibfnamefont {K.}~\bibnamefont
  {Blaum}}, \bibinfo {author} {\bibfnamefont {B.}~\bibnamefont {Cheal}},
  \bibinfo {author} {\bibfnamefont {L.}~\bibnamefont {Filippin}}, \bibinfo
  {author} {\bibfnamefont {K.T.}\ \bibnamefont {Flanagan}}, \bibinfo {author}
  {\bibfnamefont {R.F.}\ \bibnamefont {{Garcia Ruiz}}}, \bibinfo {author}
  {\bibfnamefont {W.}~\bibnamefont {Gins}}, \bibinfo {author} {\bibfnamefont
  {G.}~\bibnamefont {Gaigalas}}, \bibinfo {author} {\bibfnamefont
  {M.}~\bibnamefont {Godefroid}}, \bibinfo {author} {\bibfnamefont
  {C.}~\bibnamefont {Gorges}}, \bibinfo {author} {\bibfnamefont {L.K.}\
  \bibnamefont {Grob}}, \bibinfo {author} {\bibfnamefont {H.}~\bibnamefont
  {Heylen}}, \bibinfo {author} {\bibfnamefont {P.}~\bibnamefont {Jönsson}},
  \bibinfo {author} {\bibfnamefont {S.}~\bibnamefont {Kaufmann}}, \bibinfo
  {author} {\bibfnamefont {M.}~\bibnamefont {Kowalska}}, \bibinfo {author}
  {\bibfnamefont {J.}~\bibnamefont {Krämer}}, \bibinfo {author} {\bibfnamefont
  {S.}~\bibnamefont {Malbrunot-Ettenauer}}, \bibinfo {author} {\bibfnamefont
  {R.}~\bibnamefont {Neugart}}, \bibinfo {author} {\bibfnamefont
  {G.}~\bibnamefont {Neyens}}, \bibinfo {author} {\bibfnamefont
  {W.}~\bibnamefont {Nörtershäuser}}, \bibinfo {author} {\bibfnamefont
  {T.}~\bibnamefont {Otsuka}}, \bibinfo {author} {\bibfnamefont
  {J.}~\bibnamefont {Papuga}}, \bibinfo {author} {\bibfnamefont
  {R.}~\bibnamefont {Sánchez}}, \bibinfo {author} {\bibfnamefont
  {Y.}~\bibnamefont {Tsunoda}}, \ and\ \bibinfo {author} {\bibfnamefont {D.T.}\
  \bibnamefont {Yordanov}},\ }\bibfield  {title} {\enquote {\bibinfo {title}
  {Nuclear charge radii of $^{62-80}${Z}n and their dependence on cross-shell
  proton excitations},}\ }\href {\doibase
  https://doi.org/10.1016/j.physletb.2019.134805} {\bibfield  {journal}
  {\bibinfo  {journal} {Physics Letters B}\ }\textbf {\bibinfo {volume}
  {797}},\ \bibinfo {pages} {134805} (\bibinfo {year} {2019})}\BibitemShut
  {NoStop}%
\bibitem [{\citenamefont {Matsubara}\ \emph {et~al.}(2003)\citenamefont
  {Matsubara}, \citenamefont {Tanaka}, \citenamefont {Imajo}, \citenamefont
  {Urabe},\ and\ \citenamefont {Watanabe}}]{2003-Trap}%
  \BibitemOpen
  \bibfield  {author} {\bibinfo {author} {\bibfnamefont {K}~\bibnamefont
  {Matsubara}}, \bibinfo {author} {\bibfnamefont {U}~\bibnamefont {Tanaka}},
  \bibinfo {author} {\bibfnamefont {H}~\bibnamefont {Imajo}}, \bibinfo {author}
  {\bibfnamefont {S}~\bibnamefont {Urabe}}, \ and\ \bibinfo {author}
  {\bibfnamefont {M}~\bibnamefont {Watanabe}},\ }\bibfield  {title} {\enquote
  {\bibinfo {title} {Laser cooling and isotope-shift measurement of {Z}n$^+$
  with 202-nm ultraviolet coherent light},}\ }\href@noop {} {\bibfield
  {journal} {\bibinfo  {journal} {Applied Physics B}\ }\textbf {\bibinfo
  {volume} {76}},\ \bibinfo {pages} {209--213} (\bibinfo {year}
  {2003})}\BibitemShut {NoStop}%
\bibitem [{\citenamefont {Dzuba}\ and\ \citenamefont
  {Johnson}(2007)}]{2007-Dzuba}%
  \BibitemOpen
  \bibfield  {author} {\bibinfo {author} {\bibfnamefont {V.~A.}\ \bibnamefont
  {Dzuba}}\ and\ \bibinfo {author} {\bibfnamefont {W.~R.}\ \bibnamefont
  {Johnson}},\ }\bibfield  {title} {\enquote {\bibinfo {title} {Coupled-cluster
  single-double calculations of the relativistic energy shifts in {C IV, Na I,
  Mg II, Al III, Si IV, Ca II, and Zn II}},}\ }\href {\doibase
  10.1103/PhysRevA.76.062510} {\bibfield  {journal} {\bibinfo  {journal} {Phys.
  Rev. A}\ }\textbf {\bibinfo {volume} {76}},\ \bibinfo {pages} {062510}
  (\bibinfo {year} {2007})}\BibitemShut {NoStop}%
\bibitem [{\citenamefont {Sugar}\ and\ \citenamefont
  {Musgrove}(1995)}]{1995-Levels}%
  \BibitemOpen
  \bibfield  {author} {\bibinfo {author} {\bibfnamefont {Jack}\ \bibnamefont
  {Sugar}}\ and\ \bibinfo {author} {\bibfnamefont {Arlene}\ \bibnamefont
  {Musgrove}},\ }\bibfield  {title} {\enquote {\bibinfo {title} {{Energy Levels
  of Zinc, Zn~I through Zn~XXX}},}\ }\href@noop {} {\bibfield  {journal}
  {\bibinfo  {journal} {Journal of Physical and Chemical Reference Data}\
  }\textbf {\bibinfo {volume} {24}},\ \bibinfo {pages} {1803--1872} (\bibinfo
  {year} {1995})}\BibitemShut {NoStop}%
\bibitem [{\citenamefont {Gullberg}\ and\ \citenamefont
  {Litzén}(2000)}]{2000-Lines}%
  \BibitemOpen
  \bibfield  {author} {\bibinfo {author} {\bibfnamefont {Dag}\ \bibnamefont
  {Gullberg}}\ and\ \bibinfo {author} {\bibfnamefont {Ulf}\ \bibnamefont
  {Litzén}},\ }\bibfield  {title} {\enquote {\bibinfo {title} {{Accurately
  Measured Wavelengths of Zn I and Zn II Lines of Astrophysical Interest}},}\
  }\href {\doibase 10.1238/Physica.Regular.061a00652} {\bibfield  {journal}
  {\bibinfo  {journal} {Physica Scripta}\ }\textbf {\bibinfo {volume} {61}},\
  \bibinfo {pages} {652} (\bibinfo {year} {2000})}\BibitemShut {NoStop}%
\bibitem [{\citenamefont {Berengut}\ \emph {et~al.}(2003)\citenamefont
  {Berengut}, \citenamefont {Dzuba},\ and\ \citenamefont {Flambaum}}]{2003-CI}%
  \BibitemOpen
  \bibfield  {author} {\bibinfo {author} {\bibfnamefont {J.~C.}\ \bibnamefont
  {Berengut}}, \bibinfo {author} {\bibfnamefont {V.~A.}\ \bibnamefont {Dzuba}},
  \ and\ \bibinfo {author} {\bibfnamefont {V.~V.}\ \bibnamefont {Flambaum}},\
  }\bibfield  {title} {\enquote {\bibinfo {title} {Isotope-shift calculations
  for atoms with one valence electron},}\ }\href {\doibase
  10.1103/PhysRevA.68.022502} {\bibfield  {journal} {\bibinfo  {journal} {Phys.
  Rev. A}\ }\textbf {\bibinfo {volume} {68}},\ \bibinfo {pages} {022502}
  (\bibinfo {year} {2003})}\BibitemShut {NoStop}%
\bibitem [{\citenamefont {Stelson}\ and\ \citenamefont
  {McGowan}(1962)}]{1962-SM}%
  \BibitemOpen
  \bibfield  {author} {\bibinfo {author} {\bibfnamefont {P.H.}\ \bibnamefont
  {Stelson}}\ and\ \bibinfo {author} {\bibfnamefont {F.K.}\ \bibnamefont
  {McGowan}},\ }\bibfield  {title} {\enquote {\bibinfo {title} {{Coulomb
  excitation of the first $2^+$ state of even nuclei with $58 \leq A \leq
  82$}},}\ }\href {\doibase https://doi.org/10.1016/0029-5582(62)90368-1}
  {\bibfield  {journal} {\bibinfo  {journal} {Nuclear Physics}\ }\textbf
  {\bibinfo {volume} {32}},\ \bibinfo {pages} {652--668} (\bibinfo {year}
  {1962})}\BibitemShut {NoStop}%
\bibitem [{\citenamefont {Shera}\ \emph {et~al.}(1976)\citenamefont {Shera},
  \citenamefont {Ritter}, \citenamefont {Perkins}, \citenamefont {Rinker},
  \citenamefont {Wagner}, \citenamefont {Wohlfahrt}, \citenamefont {Fricke},\
  and\ \citenamefont {Steffen}}]{1976-Shera}%
  \BibitemOpen
  \bibfield  {author} {\bibinfo {author} {\bibfnamefont {E.~B.}\ \bibnamefont
  {Shera}}, \bibinfo {author} {\bibfnamefont {E.~T.}\ \bibnamefont {Ritter}},
  \bibinfo {author} {\bibfnamefont {R.~B.}\ \bibnamefont {Perkins}}, \bibinfo
  {author} {\bibfnamefont {G.~A.}\ \bibnamefont {Rinker}}, \bibinfo {author}
  {\bibfnamefont {L.~K.}\ \bibnamefont {Wagner}}, \bibinfo {author}
  {\bibfnamefont {H.~D.}\ \bibnamefont {Wohlfahrt}}, \bibinfo {author}
  {\bibfnamefont {G.}~\bibnamefont {Fricke}}, \ and\ \bibinfo {author}
  {\bibfnamefont {R.~M.}\ \bibnamefont {Steffen}},\ }\bibfield  {title}
  {\enquote {\bibinfo {title} {{Systematics of nuclear charge distributions in
  Fe, Co, Ni, Cu, and Zn deduced from muonic x-ray measurements}},}\ }\href
  {\doibase 10.1103/PhysRevC.14.731} {\bibfield  {journal} {\bibinfo  {journal}
  {Phys. Rev. C}\ }\textbf {\bibinfo {volume} {14}},\ \bibinfo {pages}
  {731--747} (\bibinfo {year} {1976})}\BibitemShut {NoStop}%
\bibitem [{\citenamefont {Wohlfahrt}\ \emph {et~al.}(1980)\citenamefont
  {Wohlfahrt}, \citenamefont {Schwentker}, \citenamefont {Fricke},
  \citenamefont {Andresen},\ and\ \citenamefont {Shera}}]{1980-MuE}%
  \BibitemOpen
  \bibfield  {author} {\bibinfo {author} {\bibfnamefont {H.~D.}\ \bibnamefont
  {Wohlfahrt}}, \bibinfo {author} {\bibfnamefont {O.}~\bibnamefont
  {Schwentker}}, \bibinfo {author} {\bibfnamefont {G.}~\bibnamefont {Fricke}},
  \bibinfo {author} {\bibfnamefont {H.~G.}\ \bibnamefont {Andresen}}, \ and\
  \bibinfo {author} {\bibfnamefont {E.~B.}\ \bibnamefont {Shera}},\ }\bibfield
  {title} {\enquote {\bibinfo {title} {Systematics of nuclear charge
  distributions in the mass 60 region from elastic electron scattering and
  muonic x-ray measurements},}\ }\href {\doibase 10.1103/PhysRevC.22.264}
  {\bibfield  {journal} {\bibinfo  {journal} {Phys. Rev. C}\ }\textbf {\bibinfo
  {volume} {22}},\ \bibinfo {pages} {264--283} (\bibinfo {year}
  {1980})}\BibitemShut {NoStop}%
\bibitem [{\citenamefont {Foot}\ \emph {et~al.}(1982)\citenamefont {Foot},
  \citenamefont {Stacey}, \citenamefont {Stacey}, \citenamefont {Kloch},\ and\
  \citenamefont {Le{\'s}}}]{1982-Deform}%
  \BibitemOpen
  \bibfield  {author} {\bibinfo {author} {\bibfnamefont {CJ}~\bibnamefont
  {Foot}}, \bibinfo {author} {\bibfnamefont {DN}~\bibnamefont {Stacey}},
  \bibinfo {author} {\bibfnamefont {Virginia}\ \bibnamefont {Stacey}}, \bibinfo
  {author} {\bibfnamefont {R}~\bibnamefont {Kloch}}, \ and\ \bibinfo {author}
  {\bibfnamefont {Z}~\bibnamefont {Le{\'s}}},\ }\bibfield  {title} {\enquote
  {\bibinfo {title} {Isotope effects in the nuclear charge distribution in
  zinc},}\ }\href@noop {} {\bibfield  {journal} {\bibinfo  {journal}
  {Proceedings of the Royal Society of London. A. Mathematical and Physical
  Sciences}\ }\textbf {\bibinfo {volume} {384}},\ \bibinfo {pages} {205--216}
  (\bibinfo {year} {1982})}\BibitemShut {NoStop}%
\bibitem [{\citenamefont {Ohayon}\ \emph
  {et~al.}(2022{\natexlab{a}})\citenamefont {Ohayon}, \citenamefont {Hofsäss},
  \citenamefont {Padilla-Castillo}, \citenamefont {Wright}, \citenamefont
  {Meijer}, \citenamefont {Truppe}, \citenamefont {Gibble},\ and\ \citenamefont
  {Sahoo}}]{2022-Cd}%
  \BibitemOpen
  \bibfield  {author} {\bibinfo {author} {\bibfnamefont {B}~\bibnamefont
  {Ohayon}}, \bibinfo {author} {\bibfnamefont {S}~\bibnamefont {Hofsäss}},
  \bibinfo {author} {\bibfnamefont {J~E}\ \bibnamefont {Padilla-Castillo}},
  \bibinfo {author} {\bibfnamefont {S~C}\ \bibnamefont {Wright}}, \bibinfo
  {author} {\bibfnamefont {G}~\bibnamefont {Meijer}}, \bibinfo {author}
  {\bibfnamefont {S}~\bibnamefont {Truppe}}, \bibinfo {author} {\bibfnamefont
  {K}~\bibnamefont {Gibble}}, \ and\ \bibinfo {author} {\bibfnamefont {B~K}\
  \bibnamefont {Sahoo}},\ }\bibfield  {title} {\enquote {\bibinfo {title}
  {Isotope shifts in cadmium as a sensitive probe for physics beyond the
  standard model},}\ }\href {\doibase 10.1088/1367-2630/acacbb} {\bibfield
  {journal} {\bibinfo  {journal} {New Journal of Physics}\ }\textbf {\bibinfo
  {volume} {24}},\ \bibinfo {pages} {123040} (\bibinfo {year}
  {2022}{\natexlab{a}})}\BibitemShut {NoStop}%
\bibitem [{\citenamefont {Bartlett}\ and\ \citenamefont
  {Musia\l{}}(2007)}]{bartlett}%
  \BibitemOpen
  \bibfield  {author} {\bibinfo {author} {\bibfnamefont {Rodney~J.}\
  \bibnamefont {Bartlett}}\ and\ \bibinfo {author} {\bibfnamefont {Monika}\
  \bibnamefont {Musia\l{}}},\ }\bibfield  {title} {\enquote {\bibinfo {title}
  {Coupled-cluster theory in quantum chemistry},}\ }\href {\doibase
  10.1103/RevModPhys.79.291} {\bibfield  {journal} {\bibinfo  {journal} {Rev.
  Mod. Phys.}\ }\textbf {\bibinfo {volume} {79}},\ \bibinfo {pages} {291--352}
  (\bibinfo {year} {2007})}\BibitemShut {NoStop}%
\bibitem [{\citenamefont {Crawford}(2000)}]{crawford}%
  \BibitemOpen
  \bibfield  {author} {\bibinfo {author} {\bibfnamefont {TD}~\bibnamefont
  {Crawford}},\ }\href@noop {} {\enquote {\bibinfo {title} {{HF Schaefer III in
  Reviews in Computational Chemistry, Vol. 14, KB Lipkowitz, DB Boyd, Eds}},}\
  } (\bibinfo {year} {2000})\BibitemShut {NoStop}%
\bibitem [{\citenamefont {C{\'a}rsky}\ \emph {et~al.}(2010)\citenamefont
  {C{\'a}rsky}, \citenamefont {Paldus},\ and\ \citenamefont
  {Pittner}}]{ccbook}%
  \BibitemOpen
  \bibfield  {author} {\bibinfo {author} {\bibfnamefont {Petr}\ \bibnamefont
  {C{\'a}rsky}}, \bibinfo {author} {\bibfnamefont {Josef}\ \bibnamefont
  {Paldus}}, \ and\ \bibinfo {author} {\bibfnamefont {Jir{\'\i}}\ \bibnamefont
  {Pittner}},\ }\bibfield  {title} {\enquote {\bibinfo {title} {{Recent
  progress in coupled cluster methods: Theory and applications}},}\ }\href@noop
  {} {\  (\bibinfo {year} {2010})}\BibitemShut {NoStop}%
\bibitem [{\citenamefont {Bishop}(1991)}]{bishop}%
  \BibitemOpen
  \bibfield  {author} {\bibinfo {author} {\bibfnamefont {RF}~\bibnamefont
  {Bishop}},\ }\bibfield  {title} {\enquote {\bibinfo {title} {An overview of
  coupled cluster theory and its applications in physics},}\ }\href@noop {}
  {\bibfield  {journal} {\bibinfo  {journal} {Theoretica chimica acta}\
  }\textbf {\bibinfo {volume} {80}},\ \bibinfo {pages} {95--148} (\bibinfo
  {year} {1991})}\BibitemShut {NoStop}%
\bibitem [{\citenamefont {Sahoo}(2016)}]{bijaya1}%
  \BibitemOpen
  \bibfield  {author} {\bibinfo {author} {\bibfnamefont {B.~K.}\ \bibnamefont
  {Sahoo}},\ }\bibfield  {title} {\enquote {\bibinfo {title} {{Conforming the
  measured lifetimes of the $5d{\phantom{\rule{0.16em}{0ex}}}^{2}{D}_{3/2,5/2}$
  states in Cs with theory}},}\ }\href {\doibase 10.1103/PhysRevA.93.022503}
  {\bibfield  {journal} {\bibinfo  {journal} {Phys. Rev. A}\ }\textbf {\bibinfo
  {volume} {93}},\ \bibinfo {pages} {022503} (\bibinfo {year}
  {2016})}\BibitemShut {NoStop}%
\bibitem [{\citenamefont {Yu}\ and\ \citenamefont {Sahoo}(2019)}]{bijaya2}%
  \BibitemOpen
  \bibfield  {author} {\bibinfo {author} {\bibfnamefont {Yan-mei}\ \bibnamefont
  {Yu}}\ and\ \bibinfo {author} {\bibfnamefont {B.~K.}\ \bibnamefont {Sahoo}},\
  }\bibfield  {title} {\enquote {\bibinfo {title} {{Investigating ground-state
  fine-structure properties to explore suitability of boronlike
  ${\mathrm{S}}^{11+}\ensuremath{-}{\mathrm{K}}^{14+}$ and galliumlike
  ${\mathrm{Nb}}^{10+}\ensuremath{-}{\mathrm{Ru}}^{13+}$ ions as possible
  atomic clocks}},}\ }\href {\doibase 10.1103/PhysRevA.99.022513} {\bibfield
  {journal} {\bibinfo  {journal} {Phys. Rev. A}\ }\textbf {\bibinfo {volume}
  {99}},\ \bibinfo {pages} {022513} (\bibinfo {year} {2019})}\BibitemShut
  {NoStop}%
\bibitem [{\citenamefont {Sahoo}\ and\ \citenamefont {Ohayon}(2021)}]{2021-Li}%
  \BibitemOpen
  \bibfield  {author} {\bibinfo {author} {\bibfnamefont {B.~K.}\ \bibnamefont
  {Sahoo}}\ and\ \bibinfo {author} {\bibfnamefont {B.}~\bibnamefont {Ohayon}},\
  }\bibfield  {title} {\enquote {\bibinfo {title} {{Benchmarking many-body
  approaches for the determination of isotope-shift constants: Application to
  the $\mathrm{Li}, {\mathrm{Be}}^{+}$, and ${\mathrm{Ar}}^{15+}$ isoelectronic
  systems}},}\ }\href {\doibase 10.1103/PhysRevA.103.052802} {\bibfield
  {journal} {\bibinfo  {journal} {Phys. Rev. A}\ }\textbf {\bibinfo {volume}
  {103}},\ \bibinfo {pages} {052802} (\bibinfo {year} {2021})}\BibitemShut
  {NoStop}%
\bibitem [{\citenamefont {Savukov}\ and\ \citenamefont
  {Dzuba}(2007)}]{2007-Breit}%
  \BibitemOpen
  \bibfield  {author} {\bibinfo {author} {\bibfnamefont {I.~M.}\ \bibnamefont
  {Savukov}}\ and\ \bibinfo {author} {\bibfnamefont {V.~A.}\ \bibnamefont
  {Dzuba}},\ }\href@noop {} {\enquote {\bibinfo {title} {Many-body calculations
  of relativistic energy shifts for single- and double-valence atoms},}\ }
  (\bibinfo {year} {2007}),\ \Eprint {http://arxiv.org/abs/0710.4676}
  {arXiv:0710.4676 [physics.atom-ph]} \BibitemShut {NoStop}%
\bibitem [{\citenamefont {Ohayon}\ \emph
  {et~al.}(2022{\natexlab{b}})\citenamefont {Ohayon}, \citenamefont {Ruiz},
  \citenamefont {Sun}, \citenamefont {Hagen}, \citenamefont {Papenbrock},\ and\
  \citenamefont {Sahoo}}]{2022-Na}%
  \BibitemOpen
  \bibfield  {author} {\bibinfo {author} {\bibfnamefont {B.}~\bibnamefont
  {Ohayon}}, \bibinfo {author} {\bibfnamefont {R.F.Garcia}\ \bibnamefont
  {Ruiz}}, \bibinfo {author} {\bibfnamefont {Z.~H.}\ \bibnamefont {Sun}},
  \bibinfo {author} {\bibfnamefont {G.}~\bibnamefont {Hagen}}, \bibinfo
  {author} {\bibfnamefont {T.}~\bibnamefont {Papenbrock}}, \ and\ \bibinfo
  {author} {\bibfnamefont {B.~K.}\ \bibnamefont {Sahoo}},\ }\bibfield  {title}
  {\enquote {\bibinfo {title} {{Nuclear charge radii of Na isotopes: Interplay
  of atomic and nuclear theory}},}\ }\href {\doibase
  10.1103/PhysRevC.105.L031305} {\bibfield  {journal} {\bibinfo  {journal}
  {Phys. Rev. C}\ }\textbf {\bibinfo {volume} {105}},\ \bibinfo {pages}
  {L031305} (\bibinfo {year} {2022}{\natexlab{b}})}\BibitemShut {NoStop}%
\bibitem [{\citenamefont {Dorne}\ \emph {et~al.}(2021)\citenamefont {Dorne},
  \citenamefont {Sahoo},\ and\ \citenamefont {Kastberg}}]{2021-Ca}%
  \BibitemOpen
  \bibfield  {author} {\bibinfo {author} {\bibfnamefont {Anaïs}\ \bibnamefont
  {Dorne}}, \bibinfo {author} {\bibfnamefont {Bijaya~K.}\ \bibnamefont
  {Sahoo}}, \ and\ \bibinfo {author} {\bibfnamefont {Anders}\ \bibnamefont
  {Kastberg}},\ }\bibfield  {title} {\enquote {\bibinfo {title} {{Relativistic
  Coupled-Cluster Calculations of Isotope Shifts for the Low-Lying States of Ca
  II in the Finite-Field Approach}},}\ }\href {\doibase 10.3390/atoms9020026}
  {\bibfield  {journal} {\bibinfo  {journal} {Atoms}\ }\textbf {\bibinfo
  {volume} {9}} (\bibinfo {year} {2021}),\ 10.3390/atoms9020026}\BibitemShut
  {NoStop}%
\bibitem [{\citenamefont {Xie}(2019)}]{2019-ZnThesis}%
  \BibitemOpen
  \bibfield  {author} {\bibinfo {author} {\bibfnamefont {Liang}\ \bibnamefont
  {Xie}},\ }\href@noop {} {\enquote {\bibinfo {title} {{Charge Radii of Zn and
  Ni Isotopes Measured by Collinear Laser Spectroscopy}},}\ } (\bibinfo {year}
  {2019}),\ \bibinfo {note} {the University of Manchester (United
  Kingdom)}\BibitemShut {NoStop}%
\bibitem [{\citenamefont {York}\ \emph {et~al.}(2004)\citenamefont {York},
  \citenamefont {Evensen}, \citenamefont {Martinez},\ and\ \citenamefont
  {De~Basabe~Delgado}}]{2004-York}%
  \BibitemOpen
  \bibfield  {author} {\bibinfo {author} {\bibfnamefont {Derek}\ \bibnamefont
  {York}}, \bibinfo {author} {\bibfnamefont {Norman~M}\ \bibnamefont
  {Evensen}}, \bibinfo {author} {\bibfnamefont {Margarita~L{\'o}pez}\
  \bibnamefont {Martinez}}, \ and\ \bibinfo {author} {\bibfnamefont
  {Jon{\'a}s}\ \bibnamefont {De~Basabe~Delgado}},\ }\bibfield  {title}
  {\enquote {\bibinfo {title} {{Unified equations for the slope, intercept, and
  standard errors of the best straight line}},}\ }\href {\doibase
  10.1119/1.1632486} {\bibfield  {journal} {\bibinfo  {journal} {American
  Journal of Physics}\ }\textbf {\bibinfo {volume} {72}},\ \bibinfo {pages}
  {367--375} (\bibinfo {year} {2004})}\BibitemShut {NoStop}%
\bibitem [{\citenamefont {Filippin}\ \emph {et~al.}(2017)\citenamefont
  {Filippin}, \citenamefont {Biero\ifmmode~\acute{n}\else \'{n}\fi{}},
  \citenamefont {Gaigalas}, \citenamefont {Godefroid},\ and\ \citenamefont
  {J\"onsson}}]{2017-MCDHF}%
  \BibitemOpen
  \bibfield  {author} {\bibinfo {author} {\bibfnamefont {Livio}\ \bibnamefont
  {Filippin}}, \bibinfo {author} {\bibfnamefont {Jacek}\ \bibnamefont
  {Biero\ifmmode~\acute{n}\else \'{n}\fi{}}}, \bibinfo {author} {\bibfnamefont
  {Gediminas}\ \bibnamefont {Gaigalas}}, \bibinfo {author} {\bibfnamefont
  {Michel}\ \bibnamefont {Godefroid}}, \ and\ \bibinfo {author} {\bibfnamefont
  {Per}\ \bibnamefont {J\"onsson}},\ }\bibfield  {title} {\enquote {\bibinfo
  {title} {{Multiconfiguration calculations of electronic isotope-shift factors
  in Zn i}},}\ }\href {\doibase 10.1103/PhysRevA.96.042502} {\bibfield
  {journal} {\bibinfo  {journal} {Phys. Rev. A}\ }\textbf {\bibinfo {volume}
  {96}},\ \bibinfo {pages} {042502} (\bibinfo {year} {2017})}\BibitemShut
  {NoStop}%
\bibitem [{\citenamefont {Landau}\ \emph {et~al.}(2000)\citenamefont {Landau},
  \citenamefont {Eliav}, \citenamefont {Ishikawa},\ and\ \citenamefont
  {Kaldor}}]{Kaldor}%
  \BibitemOpen
  \bibfield  {author} {\bibinfo {author} {\bibfnamefont {Arie}\ \bibnamefont
  {Landau}}, \bibinfo {author} {\bibfnamefont {Ephraim}\ \bibnamefont {Eliav}},
  \bibinfo {author} {\bibfnamefont {Yasuyuki}\ \bibnamefont {Ishikawa}}, \ and\
  \bibinfo {author} {\bibfnamefont {Uzi}\ \bibnamefont {Kaldor}},\ }\bibfield
  {title} {\enquote {\bibinfo {title} {{Intermediate Hamiltonian Fock-space
  coupled-cluster method: Excitation energies of barium and radium}},}\ }\href
  {\doibase 10.1063/1.1323258} {\bibfield  {journal} {\bibinfo  {journal} {The
  Journal of Chemical Physics}\ }\textbf {\bibinfo {volume} {113}},\ \bibinfo
  {pages} {9905--9910} (\bibinfo {year} {2000})},\ \Eprint
  {http://arxiv.org/abs/https://pubs.aip.org/aip/jcp/article-pdf/113/22/9905/10828147/9905\_1\_online.pdf}
  {https://pubs.aip.org/aip/jcp/article-pdf/113/22/9905/10828147/9905\_1\_online.pdf}
  \BibitemShut {NoStop}%
\bibitem [{\citenamefont {Chakraborty}\ \emph {et~al.}(2022)\citenamefont
  {Chakraborty}, \citenamefont {Rithvik},\ and\ \citenamefont
  {Sahoo}}]{bijaya3}%
  \BibitemOpen
  \bibfield  {author} {\bibinfo {author} {\bibfnamefont {A.}~\bibnamefont
  {Chakraborty}}, \bibinfo {author} {\bibfnamefont {S.~K.}\ \bibnamefont
  {Rithvik}}, \ and\ \bibinfo {author} {\bibfnamefont {B.~K.}\ \bibnamefont
  {Sahoo}},\ }\bibfield  {title} {\enquote {\bibinfo {title} {{Relativistic
  normal coupled-cluster theory analysis of second- and third-order electric
  polarizabilities of Zn $I$}},}\ }\href {\doibase 10.1103/PhysRevA.105.062815}
  {\bibfield  {journal} {\bibinfo  {journal} {Phys. Rev. A}\ }\textbf {\bibinfo
  {volume} {105}},\ \bibinfo {pages} {062815} (\bibinfo {year}
  {2022})}\BibitemShut {NoStop}%
\bibitem [{\citenamefont {Ford}\ and\ \citenamefont
  {Wills}(1969)}]{1969-Barret}%
  \BibitemOpen
  \bibfield  {author} {\bibinfo {author} {\bibfnamefont {Kenneth~W.}\
  \bibnamefont {Ford}}\ and\ \bibinfo {author} {\bibfnamefont {John~G.}\
  \bibnamefont {Wills}},\ }\bibfield  {title} {\enquote {\bibinfo {title}
  {Muonic atoms and the radial shape of the nuclear charge distribution},}\
  }\href {\doibase 10.1103/PhysRev.185.1429} {\bibfield  {journal} {\bibinfo
  {journal} {Phys. Rev.}\ }\textbf {\bibinfo {volume} {185}},\ \bibinfo {pages}
  {1429--1438} (\bibinfo {year} {1969})}\BibitemShut {NoStop}%
\bibitem [{\citenamefont {Angeli}\ and\ \citenamefont
  {Marinova}(2013)}]{2013-AM}%
  \BibitemOpen
  \bibfield  {author} {\bibinfo {author} {\bibfnamefont {I.}~\bibnamefont
  {Angeli}}\ and\ \bibinfo {author} {\bibfnamefont {K.P.}\ \bibnamefont
  {Marinova}},\ }\bibfield  {title} {\enquote {\bibinfo {title} {Table of
  experimental nuclear ground state charge radii: An update},}\ }\href
  {\doibase https://doi.org/10.1016/j.adt.2011.12.006} {\bibfield  {journal}
  {\bibinfo  {journal} {Atomic Data and Nuclear Data Tables}\ }\textbf
  {\bibinfo {volume} {99}},\ \bibinfo {pages} {69--95} (\bibinfo {year}
  {2013})}\BibitemShut {NoStop}%
\bibitem [{\citenamefont {Blundell}\ \emph {et~al.}(1987)\citenamefont
  {Blundell}, \citenamefont {Baird}, \citenamefont {Palmer}, \citenamefont
  {Stacey},\ and\ \citenamefont {Woodgate}}]{1987-Blund}%
  \BibitemOpen
  \bibfield  {author} {\bibinfo {author} {\bibfnamefont {S~A}\ \bibnamefont
  {Blundell}}, \bibinfo {author} {\bibfnamefont {P~E~G}\ \bibnamefont {Baird}},
  \bibinfo {author} {\bibfnamefont {C~W~P}\ \bibnamefont {Palmer}}, \bibinfo
  {author} {\bibfnamefont {D~N}\ \bibnamefont {Stacey}}, \ and\ \bibinfo
  {author} {\bibfnamefont {G~K}\ \bibnamefont {Woodgate}},\ }\bibfield  {title}
  {\enquote {\bibinfo {title} {A reformulation of the theory of field isotope
  shift in atoms},}\ }\href {\doibase 10.1088/0022-3700/20/15/015} {\bibfield
  {journal} {\bibinfo  {journal} {Journal of Physics B: Atomic and Molecular
  Physics}\ }\textbf {\bibinfo {volume} {20}},\ \bibinfo {pages} {3663}
  (\bibinfo {year} {1987})}\BibitemShut {NoStop}%
\end{thebibliography}%

\end{document}